\documentclass[12pt,preprint]{aastex}

\newcommand{\chn}{{\it Chandra}}

\usepackage{graphicx}
\usepackage{longtable}

\slugcomment{version \today: fm}

\shorttitle{\chn\ Snapshot Survey of 3C Radio Sources with z$<$0.3 II}
\shortauthors{F. Massaro et al.  2012}

\begin{document}
\title{ \chn\ Observations of 3C Radio Sources with z$<$0.3 II: \\ completing the snapshot survey}

\author{
F. Massaro\altaffilmark{1}, 
G.~R.~Tremblay\altaffilmark{2},
D.~E.~Harris\altaffilmark{3},
P.~Kharb\altaffilmark{4}
D.~Axon\altaffilmark{4,5}$^{\dagger}$,
B.~Balmaverde\altaffilmark{6}, 
S.~A.~Baum\altaffilmark{7,8}, 
A.~Capetti\altaffilmark{6},
M. Chiaberge\altaffilmark{9,10,11},
R.~Gilli\altaffilmark{12},
G.~Giovannini\altaffilmark{11,12},
P.~Grandi\altaffilmark{13},
F.~D.~Macchetto\altaffilmark{9},
C.~P.~O'Dea\altaffilmark{3,4},
G.~Risaliti\altaffilmark{14},
W.~Sparks\altaffilmark{9},
E.~Torresi\altaffilmark{13}.
}

\altaffiltext{1}{SLAC National Laboratory and Kavli Institute for Particle Astrophysics and Cosmology, 2575 Sand Hill Road, Menlo Park, CA 94025, USA}
\altaffiltext{2}{European Southern Observatory, Karl-Schwarzschild-Str. 2, 85748 Garching bei Muenchen, Germany}
\altaffiltext{3}{Smithsonian Astrophysical Observatory, 60 Garden Street, Cambridge, MA 02138, USA}
\altaffiltext{4}{Dept of Physics, Rochester Institute of Technology, Carlson Center for Imaging Science 76-3144, 84 Lomb Memorial Dr., Rochester, NY 14623, USA}
\altaffiltext{$\dagger$}{deceased}
\altaffiltext{5}{School of Mathematical and Physical Sciences, Univ. of Sussex, Brighton BN1 9QH, UK}
\altaffiltext{6}{INAF - Osservatorio Astrofisico di Torino, Strada Osservatorio 20, I-10025 Pino Torinese, Italy}
\altaffiltext{7}{Carlson Center for Imaging Science 76-3144, 84 Lomb Memorial Dr., Rochester, NY 14623, USA}
\altaffiltext{8}{Radcliffe Institute for Advanced Study, 10 Garden St. Cambridge, MA 02138, USA}
\altaffiltext{9}{Space Telescope Science Institute, 3700 San Martine Drive, Baltimore, MD 21218, USA}
\altaffiltext{10}{Center for Astrophysical Sciences, Johns Hopkins University, 3400 N. Charles Street Baltimore, MD 21218, USA}
\altaffiltext{11}{INAF - Istituto di Radioastronomia di Bologna, via Gobetti 101 40129 Bologna, Italy}
\altaffiltext{12}{INAF - Osservatorio Astronomico di Bologna, Via Ranzani 1, 40127, Bologna, Italy}
\altaffiltext{13}{INAF-IASF - Istituto di Astrofisica Spaziale e fisica cosmica di Bologna, Via P. Gobetti 101, 40129, Bologna, Italy}
\altaffiltext{14}{INAF - Osservatorio Astronomico di Arcetri, Largo E. Fermi 5, 50125, Firenze, Italy}

\begin{abstract} 
We report on the second round of \chn~ observations of the 3C snapshot
survey developed to observe the complete sample of 3C radio sources with
z$<$0.3 for 8 ksec each.  In the first paper, we illustrated the basic
data reduction and analysis procedures performed for the 30 sources of
the 3C sample observed during the \chn~ Cycle 9, while here, we
present the data for the remaining 27 sources observed during Cycle
12.  We measured the X-ray intensity of the nuclei and of any radio
hotspots and jet features with associated X-ray emission. 
{ X-ray fluxes in three
energy bands: soft, medium and hard for all the sources
analyzed are also reported}.  
For the stronger nuclei, we also applied the
standard spectral analysis which provides the best fit values of X-ray
spectral index and absorbing column density.
In addition, a detailed analysis of bright X-ray nuclei that could be affected by pileup has been performed.
X-ray emission was detected for all the nuclei of the radio sources in our sample { except for 3C\,319}. 
Amongst the current sample, there are two compact steep spectrum
{ radio sources}; two broad line radio galaxies; and one wide angle tail radio
galaxy, 3C\,89, hosted in a cluster of galaxies clearly visible in our \chn\ snapshot observation.  
In addition, we also detected soft X-ray emission arising from the galaxy cluster surrounding 3C\,196.1.  
Finally, X-ray emission from hotspots have been found in three FR\,II radio sources and, in the case of 3C\,459, 
we also report the detection of X-ray emission associated with the eastern radio lobe and
as well as that cospatial with radio jets in 3C\,29 and 3C\,402.  
\end{abstract}

\keywords{galaxies: active --- X-rays: general --- radio continuum: galaxies}

\section{Introduction}
\label{sec:intro}
In recent years several photometric and spectroscopic snapshot
surveys of 3C radio galaxies have been carried out using the Hubble
Space Telescope, approaching the statistical completeness of the radio
catalog.  A ground based spectroscopic survey for the whole sample
with the Galileo Telescope has been completed \citep{buttiglione09}
and deep ground based infrared images are also available in $K$-band.
Radio images with arcsecond resolution for the majority of the 3C
sources are available from the NRAO VLA Archive Survey (NVAS) and in
the archives of the VLA and MERLIN { observatories}. 
{ A few radio maps were made available to us by our colleagues.}
Finally, VLBA data for several of the 3C objects with $z<0.2$ have already been
obtained \citep[see e.g.,][and references therein]{giovannini01,liuzzo09}.

To extend the wavelength coverage, we embarked on a 3C snapshot survey in the
X-rays with \chn, the only X-ray facility with angular resolution
comparable to that at optical and radio frequencies.
Previous X-ray studies were mostly { based on}
samples of X-ray bright sources or objects with well-known
peculiarities instead of carefully selected and complete samples,
unbiased with respect to orientation and spectroscopic classification
as is the 3C catalog.  However, 60 sources in the 3C sample
\citep{mackay71,spinrad85}, at z$<$0.3 remained unobserved 
by \chn\ before Cycle 9, when we started
our snapshot survey.  In our first paper \citep{massaro10}, we
reported the data reduction and analysis procedures used to study the
first 30 radio sources, while in this work we present a similar
investigation of the remaining 27 sources to complete the sample at
z$<0.3$ \footnote{ This new sample includes 3C~346 that have 
been re-observed in Cycle 12 because 
during Cycle 9 its \chn\ observation was affected 
by high background \citep[see][for more details]{massaro10} .}.

{ The main aims of our snapshot survey are to detect X-ray emission
from jets and hot spots, to eventually determine their X-ray emission
processes on a firm statistical basis, and to study the nuclear
emission of the host galaxy. We also want to obtain the spectral
energy distribution (SED) of an unbiased sample of nuclear and
non-nuclear features. In a future paper, the resulting dataset will be used to test the
Fanaroff-Riley dichotomy in terms of differences in the nature of nuclear
absorption in FRI and FRII sources.  
However, it is first necessary to obtain the basic source
parameters for the newly acquired data, so this paper consists of
these data for the remaining 3C
radio galaxies at z$<$0.3.}

{ The paper} is organized as follows: a brief description of the observations and data reduction procedures is given
is \S~\ref{sec:obs}; the general and particular (i.e., single
source) results are described in \S~\ref{sec:results}.  The
conclusions and a short summary are given in \S~\ref{sec:summary}.

For our numerical results, we use cgs units unless stated otherwise
and we assume a flat cosmology with $H_0=72$ km s$^{-1}$ Mpc$^{-1}$,
$\Omega_{M}=0.27$ and $\Omega_{\Lambda}=0.73$ \citep{dunkley09}.
Spectral indices, $\alpha$, are defined by flux density,
S$_{\nu}\propto\nu^{-\alpha}$.

\section{Observations, Data Reduction, and Basic Parameters}
\label{sec:obs}
Our data reduction and data analysis procedures are described 
in Massaro et al. (2009a, 2010), here we provide only an overview.

The sources observed in this project are listed in Table~1
together with their salient parameters.  Each was observed for a
nominal 8 ksec, and the actual livetimes are given in
Table~\ref{table2}, together with the nuclear fluxes.  We used the
ACIS-S back illuminated chip in very faint mode with standard frame
times (3.2s).  All the observations had 4 chips turned on: I2, I3,
S2, and S3.  The data reduction has been performed following the
standard reduction procedure described in the \chn\ Interactive
Analysis of Observations (CIAO) threads\footnote{http://cxc.harvard.edu/ciao/guides/index.html}, using
CIAO v4.4 and the \chn\ Calibration Database (CALDB) version 4.4.8.
Level 2 event files were generated using the $acis\_process\_events$
task.  Events were filtered for grades 0,2,3,4,6.

Lightcurves for every dataset were extracted and checked for high
background intervals; none was found.

Astrometric registration was achieved by changing the appropriate keywords in the
fits header so as to align the nuclear X-ray position with that of the radio.
The celestial coordinates of the X-ray nuclei {  
were measured on displays of the event file which were regridded to a pixel size of
0.0615$^{\prime\prime}$, energy filtered for 0.5 to 7 keV, and
smoothed with a Gaussian of FWHM = 0.35$^{\prime\prime}$ or less.}
 A similar measurement was
made on the radio image and the difference in each coordinate provided
the amount of the required shift.  In most cases, the total shift
$(\sqrt{\Delta (RA)^2 + \Delta (DEC)^2}$ was of order
0.4\arcsec\ or less.

\begin{table*}  
\tiny
\caption{Source List of the \chn\ AO12 Snapshot Survey of  3C Radio Sources with z$<$0.3}\label{table1}
\begin{flushleft}
\begin{tabular}{llllrrrccccl}
\hline
3C     & Class\tablenotemark{a}  & RA (J2000)   & DEC (J2000)  & z\tablenotemark{b}       & D$_L$
          & Scale       & N$_{H,Gal}$\tablenotemark{c}    & m$_v$\tablenotemark{d} & S$_{178}$\tablenotemark{e}  & \chn\    & Obs. Date \\
          &                     & hh mm ss      & dd mm ss         &        &  (Mpc)   & (kpc/arcsec)      & (cm$^{-2}$)&     & (Jy)                 & Obs ID & yyyy-mm-dd \\ 
\hline 
\noalign{\smallskip}
~~29     &  FR I  -- LEG & 00 57 34.895 & -01 23 27.37 & 0.0448 & 193.1  & 0.858       & 3.66e20 & 14.1 & 15.1 & 12721  & 2011-05-23 \\
~~63     &  FR II  -- HEG & 02 20 54.316 & -01 56 50.72 & 0.175  & 824.7  & 2.896       & 2.47e20 & 18.5 & 19.2 & 12722  & 2010-10-27 \\ 
~~79     &  FR II -- HEG & 03 10 00.090 & +17 05 58.52 & 0.2559 & 1265.3 & 3.889       & 8.72e20 & 18.5 & 30.5 & 12723  & 2010-11-01 \\
~~89     &  WAT  --   ?  & 03 34 15.574 & -01 10 56.09 & 0.1386 & 638.1  & 2.386       & 7.02e20 & 16.0 & 20.2 & 12724  & 2010-11-03 \\
93.1   &  CSS -- HEG & 03 48 46.934 & +33 53 15.28 & 0.2430 & 1192.7 & 3.743       & 1.15e21 & 19.0 &  9.9 & 12725  & 2010-11-04 \\
130    &  FR I  --   ?  & 04 52 52.836 & +52 04 47.09 & 0.1090 & 491.9  & 1.939       & 3.66e21 & 16.5 & 15.5 & 12726  & 2010-12-11 \\
166    &  FR II -- LEG & 06 45 24.098 & +21 21 51.30 & 0.2449 & 1203.3 & 3.764       & 1.71e21 & 19.5 & 14.7 & 12727  & 2010-11-29 \\
180    &  FR II -- HEG & 07 27 04.880 & -02 04 30.33 & 0.22   & 1065.5 & 3.471       & 1.36e21 & 19.0 & 15.1 & 12728  & 2010-12-24 \\
196.1  &  FR II -- LEG & 08 15 28.10  & -03 08 28.00 & 0.198  & 946.4  & 3.197       & 5.82e20 & 17.5 & 18.6 & 12729  & 2011-02-11 \\ 
198    &   FRII    --   ?  & 08 22 31.80  & +05 57 07.90 & 0.0815 & 360.8  & 1.496       & 2.24e20 & 16.8 &  9.7 & 12730  & 2011-01-07 \\ 
223    &  FR II -- HEG & 09 39 52.755 & +35 53 58.86 & 0.1368 & 629.0  & 2.360       & 1.04e20 & 17.1 & 14.7 & 12731  & 2012-01-07 \\ 
234    &  FR II -- HEG & 10 01 49.526 & +28 47 08.87 & 0.1848 & 876.2  & 3.026       & 1.76e20 & 17.3 & 31.4 & 12732  & 2011-01-19 \\ 
258    &  CSS --  ?   & 11 24 43.881 & +19 19 29.50 & 0.165?  & 772.7  & 2.760       & 1.51e20 & 19.5 &  9.7 & 12733  & 2010-11-01 \\ 
284    &  FR II -- HEG & 13 11 04.666 & +27 28 07.15 & 0.2394 & 1172.6 & 3.701       & 9.75e19 & 18.0 & 11.3 & 12735  & 2010-11-17 \\
314.1  &  FR I  -- LEG & 15 10 27.064 & +70 46 07.37 & 0.1197 & 544.2  & 2.104       & 1.93e20 & 17.0 & 10.6 & 12736  & 2012-01-02 \\ 
319    &  FR II --  ?   & 15 24 05.640 & +54 28 18.40 & 0.192  & 914.3  & 3.120       & 1.16e20 & 17.0 & 10.6 & 12736  & 2010-10-25 \\ 
357    &  FR II -- LEG & 17 28 20.109 & +31 46 02.58 & 0.1662 & 778.9  & 2.776       & 3.07e20 & 15.5 &  9.7 & 12738  & 2010-10-31 \\
379.1  &  FR II -- HEG & 18 24 32.976 & +74 20 59.00 & 0.256  & 1265.8 & 3.890       & 5.43e20 & 18.0 &  7.4 & 12739  & 2011-04-04 \\
402    &  FR I  --  ?   & 19 41 45.899 & +50 35 45.86 & 0.0239 & 101.4  & 0.469       & 1.07e21 & 14.0 & 10.1 & 12740  & 2011-07-12 \\
403.1  &   FRII  -- LEG & 19 52 30.50  & -01 17 18.00 & 0.0554 & 240.7  & 1.048       & 1.19e21 & 16.0 & 13.5 & 12741  & 2010-11-27 \\ 
410    &  FR II -- BLO & 20 20 06.60  & +29 42 14.20 & 0.2485 & 1223.5 & 3.805       & 4.35e21 & 19.5 & 34.6 & 12742  & 2011-09-24 \\ 
424    &  FR I  -- LEG & 20 48 12.087 & +07 01 17.17 & 0.127  & 580.2  & 2.215       & 7.05e20 & 18.0 & 14.6 & 12743  & 2011-04-15 \\ 
430    &  FR II -- LEG & 21 18 19.094 & +60 48 07.77 & 0.0541 & 234.8  & 1.025       & 3.31e21 & 15.0 & 33.7 & 12744  & 2011-11-14 \\ 
436    &  FR II -- HEG & 21 44 11.727 & +28 10 18.92 & 0.2145 & 1035.5 & 3.403       & 6.43e20 & 18.2 & 17.8 & 12745  & 2011-05-27 \\ 
456    &  FR II -- HEG & 23 12 28.076 & +09 19 26.39 & 0.2330 & 1137.0 & 3.626       & 3.70e20 & 18.5 & 10.6 & 12746  & 2011-01-17 \\
458    &  FR II -- HEG & 23 12 52.083 & +05 16 49.77 & 0.289  & 1455.2 & 4.246       & 5.85e20 & 20.0 & 14.5 & 12747  & 2010-10-10 \\
459    &  FR II -- BLO & 23 16 35.30  & +04 05 18.30 & 0.2199 & 1064.9 & 3.469       & 5.24e20 & 17.6 & 25.6 & 12734  & 2011-10-13 \\ 
\noalign{\smallskip}
\hline
\end{tabular}\\
\end{flushleft}
(a) The 'class' column contains both a radio descriptor (Fanaroff-Riley class I or II), Compact Steep Spectrum (CSS), Wide Angle Tail (WAT) radio galaxy and 
the optical spectroscopic designation, LEG, ``Low Excitation Galaxy'', HEG, ``High Excitation Galaxy'', and BLO, ``Broad Line Object''.
The symbol ``?" indicates those optical classification that are uncertain or not reported in the literature.\\
(b) Redshift measurements are taken form Chiaberge et al. (2002), Floyd et al. (2008), Buttiglione et al. (2009).\\
(c) { Galactic Neutral hydrogen column densities N$_{H,Gal}$} are taken form Kalberla et al. (2005).\\
(d) $m_v$ is the visual magnitude (Spinrad et al. 1985).\\
(e) S$_{178}$ is the flux density at 178 MHz, taken from Spinrad et al. (1985).\\
\end{table*}

\subsection{Fluxmaps}
\label{sec:fluxmaps}
We created 3 different fluxmaps (soft, medium, hard, in the ranges 0.5
-- 1, 1 -- 2, 2 -- 7 keV, respectively) by filtering the event file with
the appropriate energy range and dividing the data with
monochromatic exposure maps (with nominal energies of soft=0.8keV,
medium=1.4keV, and hard=4keV).  The exposure maps and the fluxmaps
were regridded to a common pixel size which was usually 1/4 the size
of a native ACIS pixel (native=0.492\arcsec). 
In fact to achieve the angular resolution of the Chandra mirrors, 
we need to avoid the undersampling imposed by the ACIS pixel size. 
{ Since the location of each event is a real number and not an integer
value denoting only the center of a pixel location, and since Chandra
routinely dithers on the sky, avoiding undersampling can be achieved
by regridding to obtain pixel sizes of 0.123$^{\prime\prime}$ or
smaller.} For sources of large angular extent we used 1/2 or no regridding.

To obtain maps with brightness units of ergs~cm$^{-2}$~s$^{-1}$~pixel$^{-1}$, we 
multiplied each event by the nominal energy of its respective
band \citep[see also][]{massaro09b}.

To measure observed fluxes for any feature, we construct an
appropriate region (usually circular) and two adjacent background
regions of the same size.  The two background regions were chosen so
as to avoid contaminating X-ray emission (and also radio emission) and
permitted us to sample both sides of jet features, two areas close to
hotspots, and avoid contamination from weak emission surrounding the
nuclei of the galaxies \citep[e.g.,][]{massaro11}.

{ The total energy in any particular band for any particular region is measured with
funtools\footnote{http://www.cfa.harvard.edu/$\sim$john/funtools}, 
and with our choice of units for the band map, comes out in
cgs units.  The use of the ``nominal energy" is solely to get the
correct units, and each event in the region is assumed to have $h*\nu$
ergs where $\nu$ is assigned a nominal value for each band.  The
actual energy in the band is the sum of all events within the region,
each with its observed energy.  By applying a correction factor of
$E(average)/E(nominal)$ we recover the actual total energy for that
particular region since to derive E(average) the actual E values are used.} 
This correction ranged from a few to 15\%.  A one $\sigma$ error is assigned based on
the usual $\sqrt{number-of-counts}$ in the on and background regions.
Fluxes for the nuclei are given in Table~2.

\begin{table*} 
\tiny
\caption{Nuclear X-ray Fluxes in units of 10$^{-15}$erg~cm$^{-2}$s$^{-1}$}\label{table2}
\begin{flushleft}
\begin{tabular}{llrrrrrrrrrr}
\hline
3C     & LivTim\tablenotemark{a}&  Net\tablenotemark{b}& Ext. Ratio\tablenotemark{c}& f(soft)          & f(medium)& f(hard)    & f(total)        & HR & N$_H^d$ & L$_X$ \\ 
          & (ksec)                                 & (cnts)                             &                                                   & 0.5-1~keV & 1-2~keV   & 2-7~keV & 0.5-7~keV &                                      & (10$^{22}$cm$^{-2}$)  & (10$^{42}$erg~s$^{-1}$) \\
\hline 
\noalign{\smallskip}
  ~~29   & 7.95    & 51(7)     &  0.52(0.11)   & 2.9(1.1)   & 11.9(2.3)  & 29.1(6.7)  & 43.9(7.2)  &  0.42(0.19) & $<$1.81      &    0.19(0.03)\\ 
  ~~63   & 7.95    & 513(22)   &  0.91(0.06)   & 9.0(2.1)   & 101(7)     & 496(30)    & 606(31)    &  0.66(0.06) & 2.19 - 5.18  &    49.3(2.5)\\ 
  ~~79   & 7.69    & 104(10)   &  0.91(0.13)   & 3.4(1.2)   & 3.7(1.3)   & 200(21)    & 207(21)    &  0.96(0.15) & 5.24 - 24.5  &    39.7(4.1)\\ 
  ~93.1  & 7.69    & 101(10)   &  0.97(0.13)   & 10.0(2.3)  & 18.6(2.8)  & 56.1(9.2)  & 98.7(9.9)  &  0.50(0.14) & 1.09 - 4.85  &    14.4(1.7)\\ 
  130    & 7.95    & 28(5)     &  0.54(0.20)   & 2.4(1.1)   & 7.2(1.9)   & 12.7(4.8)  & 22.3(5.3)  &  0.28(0.27) & $<$3.27      &    0.64(0.15)\\
  166    & 7.95    & 432(20)   &  0.96(0.06)   & 37.3(4.2)  & 94.4(6.5)  & 238(20)    & 370(22)    &  0.43(0.07) & 1.07 - 3.63  &    64.0(3.7)\\
  180    & 7.95    & 26(5)     &  0.93(0.27)   & 1.90(0.85) & 2.1(1.0)   & 40.3(9.8)  & 44.3(9.9)  &  0.90(0.31) & 2.35 - 24.5  &    6.0(1.3)\\
  198    & 7.95    & 5(2)      &  0.83(0.57)   & --         & 0.74(0.53) & --         & 0.74(0.53) &   --        & --           &    0.04(0.02)\\
  223    & 7.95    & 113(11)   &  0.95(0.13)   & 5.9(1.6)   & 15.6(2.7)  & 130(16)    & 151(16)    &  0.79(0.14) & 2.34 - 8.79  &    7.17(0.78)\\
  234    & 7.95    & 311(17)   &  0.95(0.07)   & 135(19)    & 81(11)     & 433(31)    & 649(37)    &  0.69(0.08) & 2.24 - 5.74  &    59.9(3.5)\\ 
  258    & 7.95    & 6(2)      &  0.85(0.99)   & --         & 1.40(0.81) & 5.4(3.9)   & 7.3(4.0)   &  0.68(0.52) & $<$24.4      &    0.52(0.29)\\
  284    & 7.95    & 16(4)     &  1.00(0.40)   & 2.6(1.1)   & 3.0(1.1)   & 5.4(3.1)   & 11.1(3.5)  &  0.29(0.41) & $<$3.29      &    1.83(0.58)\\
  314.1  & 7.95    & 14(4)     &  1.00(0.40)   & 1.81(0.91) & 3.1(1.1)   & 6.1(3.6)   & 11.0(3.9)  &  0.33(0.43) & $<$5.11      &    0.39(0.14)\\
  357    & 7.95    & 129(11)   &  0.95(0.12)   & 2.5(1.0)   & 18.4(3.0)  & 145(16)    & 166(16)    &  0.77(0.13) & 2.45 - 8.68  &    12.0(1.2)\\
  379.1  & 7.95    & 30(5)     &  0.97(0.25)   & --         & 2.01(0.90) & 54(11)     & 56(11)     &  0.93(0.27) & 3.25 - 24.5  &    10.8(2.1)\\
  402    & 7.95    & 167(13)   &  0.83(0.09)   & 21.7(3.1)  & 35.9(3.9)  & 55.2(9.1)  & 113(10)    &  0.21(0.11) & $<$1.00      &    0.14(0.01)\\
  403.1  & 7.95    & 6(2)      &  1.00(0.74)   & --         & 1.25(0.72) & 2.5(1.8)   & 4.3(2.0)   &  0.33(0.53) & $<$6.70      &    0.03(0.01)\\
  410    & 7.95    & 1122(34)  &  0.89(0.04)   & 24.6(3.5)  & 193(10)    & 1220(47)   & 1437(48)   &  0.73(0.04) & 3.23 - 6.41  &    257(9)\\
  424    & 7.95    & 52(7)     &  0.67(0.13)   & 1.24(0.89) & 9.5(2.2)   & 58(10)     & 68(10)     &  0.72(0.19) & 1.56 - 8.03  &    2.76(0.42)\\  
  430    & 7.95    & 8(3)      &  0.62(0.41)   & --         & 1.23(0.71) & 10.6(4.7)  & 11.8(4.8)  &  0.79(0.52) & $<$24.1      &    0.08(0.03)\\
  436    & 7.95    & 39(6)     &  0.76(0.18)   & 1.00(0.58) & 0.91(0.64) & 80(14)     & 81(14)     &  0.98(0.24) & 3.75 - 24.5  &    10.5(1.8)\\
  456    & 7.95    & 328(18)   &  0.92(0.07)   & 3.1(1.2)   & 15.9(2.8)  & 545(32)    & 564(32)    &  0.94(0.08) & 5.24 - 24.4  &    87.2(5.0)\\
  458    & 7.95    & 37(6)     &  1.00(0.25)   & --         & 2.15(0.97) & 72(13)     & 74(13)     &  0.94(0.24) & 3.67 - 24.5  &    18.7(3.2)\\
  459    & 7.95    & 100(10)   &  0.87(0.12)   & 13.8(2.4)  & 16.7(2.7)  & 58(10)     & 88(11)     &  0.55(0.16) & 1.18 - 5.63  &    12.0(1.5)\\
\noalign{\smallskip}
\hline
\end{tabular}\\
\end{flushleft}
\tablecomments{Values in parentheses are 1$\sigma$ uncertainties.}

\tablenotemark{a}{LivTim is the live time} 

\tablenotemark{b}{Net is the net counts within a circle of radius=
2\arcsec.}

\tablenotemark{c}{Ext. Ratio (``Extent Ratio'') is the ratio of the
net counts in the r\,=\,2\arcsec\ circle to the net counts in the
r\,=\,10\arcsec\ circle.  Values significantly less than 0.9
indicate the presence of extended emission around the nuclear
component. A more detailed analysis of the radio sources 
with extended emission around their nuclei
will be presented in Balmaverde et al. (2012).}  

\tablenotemark{d}{As per the discussion in the text, we calculate the
value of N$_H$ required to produce the observed $HR$ values.  
The uncertainty given here is indicative only: it is the range
of N$_H$ covered by the uncertainty in the $HR$ and allowing
$\alpha_X$ to range from 0.5 to 1.5.  Obviously there may be some
sources with intrinsic spectral indices outside of this range.}

\end{table*}

\begin{table}
\tiny
\caption{Estimates of intrinsic N$_H$ for the AO9 sources that are not affected by pileup}
\begin{flushleft}
\begin{tabular}{lrrr}
\hline
3C     & redshift & HR & N$_H$ \\ 
       &          &    & (10$^{22}$cm$^{-2}$)\\
\hline 
\noalign{\smallskip}
  20    & 0.174  & 0.99(0.12) & 5.42 - 24.4\\
  52    & 0.2854 & 0.86(0.47) & 1.29 - 24.5\\
  61.1  & 0.1878 & 0.84(0.21) & 2.39 - 24.4\\
  76.1  & 0.0324 & 0.78(0.29) & 1.13 - 21.5\\
  105   & 0.089  & 0.98(0.08) & 3.03 - 23.6\\
  132   & 0.214  & 0.94(0.21) & 3.49 - 24.5\\
  135   & 0.1273 & 0.80(0.26) & 1.63 - 24.2\\
  165   & 0.2957 & 0.79(0.24) & 2.27 - 24.5\\
  171   & 0.2384 & 0.96(0.11) & 5.82 - 24.5\\
  213.1 & 0.1937 & 0.22(0.19) & $<$ 2.82\\
  223.1 & 0.1075 & 0.97(0.14) & 4.00 - 24.2\\
  293   & 0.045  & 0.97(0.10) & 3.80 - 21.5\\
  300   & 0.27   & 0.37(0.12) & $<$ 3.92\\
  303.1 & 0.267  & 0.68(0.47) & $<$ 24.5\\
  305   & 0.0416 & 0.14(0.22) & $<$ 1.76\\
  315   & 0.1083 & 1.00(0.71) & $<$ 24.2\\
  349   & 0.205  & 0.68(0.09) & 2.32 - 6.57\\
  381   & 0.1605 & 0.96(0.09) & 5.14 - 24.4\\
  436   & 0.2145 & 0.94(0.23) & 2.25 - 24.5\\
  460   & 0.268  & 0.88(0.24) & 2.93 - 24.5\\
\noalign{\smallskip}
\hline
\end{tabular}\\
\end{flushleft}
\tablecomments{Values in parentheses are 1$\sigma$ uncertainties.
As for Table~2, we calculate the
value of N$_H$ required to produce the observed $HR$ values.  
The uncertainty given here is indicative only: it is the range
of N$_H$ covered by the uncertainty in the $HR$ and allowing
$\alpha_X$ to range from 0.5 to 1.5 (see Section 3.2 for more details).}

\end{table}

\subsection{X-ray Spectral Analysis of the stronger nuclei}
\label{sec:spectra}
We have performed { an X-ray} spectral analysis for those nuclear point sources
containing 250 or more counts, so as to quantify their { X-ray}
spectral indices $\alpha_X$, the presence or absence of significant intrinsic
absorption, and the role played by mild pileup in artificially
hardening the spectrum.

The spectral data were extracted from a 1\arcsec.5 aperture using the
{\sc ciao} 4.4 routine \texttt{specextract}, thereby automating the creation
of count-weighted response matrices. The background-subtracted spectra
were then filtered in energy between 0.3-7 keV, binned using a
30 count threshold, and fit with absorbed power-law models using
iterative $\chi^2$ minimization techniques with {\sc xspec} version 12.6
\citep{arnaud96}.

Two multiplicative models are fit to each source: (1) a simple
redshifted powerlaw with Galactic and intrinsic photoelectric
absorption components\\
(\texttt{phabs}$\times$\texttt{zphabs}$\times$\texttt{zpowerlaw} in
{\sc xspec} syntax), \\ and (2) the same model with an additional pileup
component,\\
(\texttt{pileup}$\times$\texttt{phabs}$\times$\texttt{zphabs}$\times$\texttt{zpowerlaw}),
using the {\sc xspec} implementation of the \chn\ pileup
model described by Davis (2001).

Prior to fitting, the Milky Way hydrogen column density and the source
redshift were fixed.  The two main variable parameters, namely the
intrinsic absorption N$_H\left(z\right)$ and { X-ray} photon index $\Gamma$ 
were allowed to
vary in a first pass fit, but subsequently stepped through a range of
possible physical values to explore the parameter space,
determine 90\% confidence intervals, and quantify the degree to which
N$_H\left(z\right)$ and { $\Gamma$} are degenerate.

Monte Carlo Markov Chains were created to further aid our
understanding of these behaviors. 
We present our results in Table~4.  We have
also explored the possible effect of pileup on our sources, which we
discuss in the section below.

{ We also note that those sources with inverted best-fit spectral
indices ($\Gamma < 1$ or $\alpha < 0$) are likely to be unphysical for the
power-law plus absorption model.  Inverted spectra for the
energy band 0.5-7keV can result from Compton Thick models.}
3C 79, 3C 105, 3C 234 and 3C 332 are 
not included in this table as no constraints can be found using an absorbed power law model. 
A detailed analysis of the X-ray spectrum of 3C 234 can be found in Piconcelli et al. (2008)
{ while the X-ray spectrum of 3C 105 is presented in Massaro et al. (2010).} 
Those sources whose count rates proved insufficient for $\chi^2$ minimization 
techniques were instead fit using Cash (1979) statistics.

\begin{table*}
\tiny
\label{table3}
\caption{Spectral Analysis of Brighter Nuclei}
\begin{flushleft}
\begin{tabular}{lccc|ccccc}
\hline
             &        &                         & $N_{H,Gal}$                                                       &                    & $N_H$ (int.)                                                        & Pileup G.M.&                   \\
Source  & $z$ & FR/Opt.~class & $\left( \times 10^{20} ~\mathrm{cm}^{-2} \right)$ & $\Gamma$& $\left( \times 10^{20} ~\mathrm{cm}^{-2} \right)$& (\%)            & $\chi^2/$dof\\
(1) & (2) & (3) & (4) & (5) & (6) & (7) & (8) \\
\hline
\hline
3C\,17*     &  0.2197  &  II / BLO &  2.86  & $1.75^{+0.12}_{-0.11}$  & $[0.0]^{+3.4}_{\downarrow}$     & (0.94$^\dagger$) & 22.6/31 \\
3C\,18*     &  0.188   &  II / BLO &  5.33  & $1.12^{+0.17}_{-0.11}$  & $[0.0]^{+9.0}_{\downarrow}$     & (1.0$^\dagger$)  & 40.6/28 \\
3C\,33.1*   &  0.181   &  II / BLO & 20.0   & 1.47$\pm$0.2            & 469$\pm$135                     & (1.0)            & 19.3/18 \\ 
3C\,63      &  0.175   &  II / HEG &  2.47  & $(1.36^\dagger)$        & $71.0^{+28.0}_{-26.0}$          & (1.0$^\dagger$)  &  3.6/13 \\
3C\,133*    &  0.278   &  II / HEG & 25.4   & 2.63$\pm$0.2            & 116$\pm$45                      & (1.0)            & 18.3/15 \\ 
3C\,166     &  0.245   &  II / LEG & 17.1   & $1.65^{+0.27}_{-0.22}$  & $[0.0]^{+12.3}_{\downarrow}$    & (1.0)            &  9.41/9 \\
3C\,184.1*  &  0.118   &  II / BLO &  3.15  & 1.80$\pm$0.5            & 358$\pm$165                     & (1.0)            & 12/13   \\ 
3C\,197.1*  &  0.130   &  II / BLO &  4.20  & 1.64$\pm$0.13           & $[0.0]^{+7.4}_{\downarrow}$     & (1.0)            & 24.5/24 \\ 
3C\,287.1*  &  0.2156  &  II / BLO &  1.63  & $1.43^{+0.10}_{-0.10}$  & $[0.0]^{\uparrow}_{\downarrow}$ & (1.0$^\dagger$)  & 24.6/32 \\
3C\,323.1*  &  0.2643  &  II / BLO &  3.79  & $0.406^{+0.15}_{-0.15}$ & $[0.0]^{\uparrow}_{\downarrow}$ & (1.0)            & 16.6/13 \\
3C\,410     &  0.249   &  II / BLO & 43.5   & $(1.78^\dagger)$        & $66.7^{+21.0}_{-19.1}$          & (1.0)            & 30.5/32 \\
3C\,456     &  0.233   &  II / HEG &  3.70  & $(1.89^\dagger)$        & $867.9^{+174}_{-150}$           & (1.0$^\dagger$)  &  3.8/7  \\
%
\noalign{\smallskip}
\hline
\end{tabular}\\
\end{flushleft}
{ The model used for the data fitting procedure is \texttt{pileup} $\times$ \texttt{phabs} $\times$ \texttt{zphabs}  $\times$ \texttt{zpowerlaw} in {\sc xspec} syntax. 
The number of counts per bin is shown in parentheses. Spectra with 30 count bins.
For these fits, we show in square brackets the corresponding range of best-fit $N_H$ as the photon index $\Gamma$ is stepped through values of 1.0 to 3.0\\
Col. (1): source name (an asterisk indicates an AO09 source adversely affected by pileup, warranting a re-analysis).\\
Col. (2): source redshift (this parameter was frozen in all spectral fits).\\
Col. (3): Fanaroff-Riley radio classification \citep{fanaroff74} and optical emission line classification (HEG: ``high excitation galaxy''; LEG: ``low excitation galaxy''; BLO: ``broad-line object'', classifications \citep{buttiglione09}).\\
Col. (4): Galactic absorption column density in units of $\times 10^{20}$ cm$^{-2}$ (Kalberla et al. 2005; this parameter was frozen in all spectral fits).\\
Col. (5): photon index (note that $\alpha_X = \Gamma -1$. A common value for unabsorbed AGN is $\Gamma=1.9$ or $\alpha_X=0.9$.).\\
Col. (6): intrinsic absorbing column density in units of 10$^{20}$ cm$^{-2}$.\\
{ - Parentheses `()' indicates that the value was frozen for the listed fit. \\
- A dagger symbol means that the parameter was initially left free in a first-pass fit, then subsequently frozen to its original best fit value. \\
- A bracket around a number means that the parameter is not constrained. We also show a number in brackets with an upper limit listed.
A down arrow indicates that no lower bound is found. \\
- A bracketed range (e.g., [1.0 ---> 3.0]) indicates that a series of fits were computed using a grid of values ranging between the limits listed in brackets.
For the column density $N_H$, we list the corresponding best-fit values as we step through this range of $\Gamma$. \\}
Col. (7): grade migration parameter from \texttt{jdpileup}.\\
Col. (8): $\chi^2$ / degrees of freedom.\\}
\end{table*}

\subsection{The effects of mild pileup}\label{sec:pileup}
In our previous data paper on the AO9 sample of 3C sources
\citep{massaro10}, we erroneously neglected the effects of mild pileup in
the ACIS CCD's (but see the note added in proof, {\it ibid}).  In this
section we describe our diagnostic for pileup and suggest a threshold
countrate, above which spectral analysis and our use of hardness
ratios to estimate the excess absorption (the column density of N$_H$
ascribed to material at the source) are suspect.  The application of
{ the 'jdpileup model' in Sherpa \footnote{\underline{http://cxc.harvard.edu/sherpa/}} \citep{freeman01} and in 
{\sc xspec} \citep{arnaud96} for our data is discussed in
$\S$\ref{sec:spectra}.}

\begin{figure}
\includegraphics[scale=0.45]{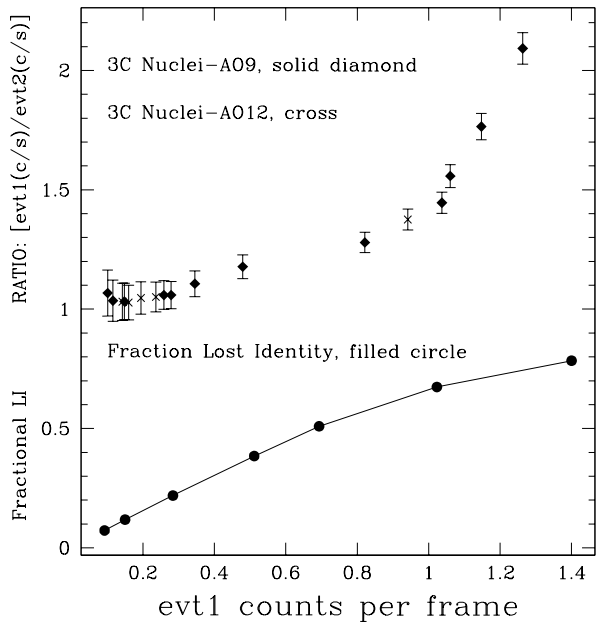}
\caption{As a function of the total event 1 count rate (counts per
  frame) we show the ratio of evt1 countrate to evt2 countrate for the
  brighter nuclei from the \chn\ 3C snapshot survey.  These data are
  the values $\geq$1 (with 1$\sigma$ statistical error bars).  The
  plotted values below 1 are the calculated fractions of incoming
  photons which loose their identity by ending up in piled
  events.\label{fig:pileup}}
\end{figure}

As described in $\S$3.2 of Harris et al. (2011), we have verified that the
standard grade filtering that is applied to all ACIS data in pipeline
processing means that a fraction of piled events are rejected in
passing from the event 1 files to event 2.  This is caused by 'grade
migration', an inevitable consequence of pielup \citep{davis01}.  Thus
we choose the ratio of countrates (evt1/evt2) as a robust diagnostic
of pileup.  In Figure~\ref{fig:pileup} we demonstrate this for our
sources by plotting the ratio against the event 1 counts per frame.
{ We have also checked weaker sources and
verified that their evt1 and evt2 countrates are equal within the errors.}

We have used PIMMS to elucidate what this means in terms of pileup
fraction.  While PIMMS does not accommodate grade migration, by using
the evt1 countrate, we are able to derive all necessary quantities.
For a series of input c/s to PIMMS, with a power law spectrum of
$\alpha_X$=1 and no absorption, PIMMS returns the fractional countrate,
Fp which is the fraction of piled events to total (i.e. evt1) events.
The 'output' countrate of PIMMS is the rate of unpiled events.  Thus
the total countrate (our observed evt1 c/s) is given by:
\begin{equation}
evt1(c/s) = \frac{out(c/s)}{(1-Fp)} 
\end{equation}
In judging `how bad things can get', we prefer a new parameter,
'F$_{li}$', rather than the reported fraction of piled to total
events.  We define F$_{li}$ as the fraction of incoming {\it photons}
which loose their identity to the total number of incoming photons.
So F$_{li}$ tells us what fraction of the total photons end up in
piled events:
\begin{equation}
F_{li} = \frac{IN - OUT}{IN}
\end{equation}
where IN is the input c/s to PIMMS, and OUT is the count rate PIMMS
reports, i.e. the unpiled rate.  F$_{li}$ is plotted in
Figure~\ref{fig:pileup}.

The obvious question then is, "As counts/frame increase, when would we
start getting significantly wrong spectral parameters from Sherpa?".
{ The answer to this question is not unique: it depends on the spectral
distribution of the source and on the accuracy of our estimates of
spectral parameters (and hence on the number of events available for
spectral analysis).} 
For our purposes, we subjectively assign 0.2 cnts/frame (where more than 10\%
of incoming photons end up in piled events) as the threshold above
which, we should be circumspect of spectral parameters.  For our data
with 3.2s frame time and an exposure time of 8ks, this corresponds to
0.06 c/s and 480 total counts.  This criterion applies to { 1} source in
the AO12 sample and 9 from the AO9 sample.

\section{Results}
\label{sec:results}

\subsection{General}
\label{sec:general}
X-ray emission was detected for all the nuclei in the sample except
for 3C\,319, an FR II radio galaxy of small angular size.  The observed nuclear
fluxes are presented in Table~2 in the soft, medium, and
hard bands together with the X-ray luminosity and the X-ray hardness
ratio $HR$ \citep[e.g.,][]{massaro10,massaro11}.  This has been
evaluated according to the simple relation: $(H-M)/(H+M)$, where $H$
and $M$ are the X-ray fluxes in the hard and the medium bands,
respectively.  In Table~2 the uncertainties on the
observed values of $HR$, have been derived from the errors on the flux
values.
{ The $HR$ values have not been computed using the soft band because it is the band most
affected by absorption.  Thus for large column densities, there are
very few counts (or none) in the soft band.  Values of HR are not
provided for 3C 319 (no detection of nucleus) and the two sources, 3C89
and 3C196.1 which lie in clusters of galaxies, making it difficult to
separate nuclear emission from the relatively bright X-ray emission
from the cluster gas.}

In addition, as performed for the previous subset of 3C sources
observed during the \chn\ Cycle 9, we measured the net number of
counts within circular regions of { radii} 2\arcsec\ and
10\arcsec, both centered on the nucleus of each source.
In Table~2 we give the r=2\arcsec\ result
together with the ratio of r2/r10, a diagnostic for the presence of
extended emission around the nucleus. The only exceptions are 3C\,89
and 3C\,196.1, that lie in X-ray detected galaxy clusters, as shown in
Figs.~\ref{fig:3c89a},~\ref{fig:3c89b},~\ref{fig:3c196.1a}~and~\ref{fig:3c196.1b},
as well as 3C\,319 that is undetected.  This ratio should be close to
unity for an unresolved source: the on-axis encircled energy for
r=2\arcsec\ is $\approx$0.97 so we expect only a small
increase between r=2\arcsec\ and r=10\arcsec\ for an
unresolved source.

Amongst the X-ray detected sources we found two compact steep spectrum
(CSS) radio sources: 3C\,93.1 and 3C\,258, two broad line radio
galaxies: 3C\,410 and 3C\,459, and one wide angle tail (WAT) radio
galaxy, 3C\,89, hosted in a cluster of galaxies, with the largest known
X-ray cavity \citep{sun12}.  In addition, we detected soft X-ray
emission arising from the galaxy cluster surrounding the FR\, II radio
galaxy 3C\,196.1.  For 3 of our sources we have hotspot detections in
our \chn\ images, with confidence levels between { 2.0$\sigma$ and 3.2$\sigma$}.
In addition, in the case of 3C\,459 we also detect the X-ray emission
arising from the eastern radio lobe.  Finally we detected X-ray
emission cospatial with radio jets in 3C\,29 (see
Figure~\ref{fig:3c29app}) and 3C\,402.  Fluxes for both jet and hotspot
structures found in our sample are reported in Table~5.
{ In the same table we also provide the confidence level of each detection
evaluated adopting a Poisson distribution.}

\begin{table*}  
\tiny
\caption{Radio components with X-ray Detections}\label{tab:extend}
\begin{center}
\begin{tabular}{llllllllll}
\hline
3C                     &Component\tablenotemark{a}& Radius\tablenotemark{b}&counts\,(bkg)\tablenotemark{c}& Detection &   
f$_{0.5-1~keV}$&f$_{1-2~keV}$                        &f$_{2-7~keV}$                &f$_{0.5-7~keV}$                       &L$_X$\\ 
                          &                                               & (arcsec)                         &                                                 & Significance\tablenotemark{d} &
(cgs)                  &(cgs)                                       &(cgs)                               &(cgs)                                         &10$^{41}$erg~s$^{-1}$\\
\hline 
\noalign{\smallskip}
 ~~29  & $k$ - s\,2.7 & 1.0 &  8(1) & 4.7$\sigma$ & 1.39(0.80) & 1.67(0.84) & --         &  3.1(1.2) & 0.014(0.005)\\        
  234  & $h$ - w\,45  & 2.0 &  5(1) & 3.2$\sigma$ & --         & 0.86(0.62) & 3.2(3.1)   &  4.1(3.3) & 0.38(0.30)\\
  402  & $k$ - s\,1.7 & 0.8 & 18(7) & 3.6$\sigma$ & 4.3(1.9)   & 1.25(0.86) & 1.58(1.54) &  7.1(2.6) & 0.009(0.003)\\
  436  & $h$ - s\,43  & 1.5 &  4(1) & 2.7$\sigma$ & --         & 0.92(0.65) & --         &  3.2(2.0) & 0.41(0.26)\\
  458  & $h$ - e\,104 & 2.0 &  3(1) & 2.0$\sigma$ & --         & 0.78(0.55) & --         &  2.2(1.5) & 0.57(0.39)\\
  459  & $l$ - e\,1.3 & 0.8 & 24(13)& 2.9$\sigma$ & 4.2(1.8)   & --         & --         &  4.2(1.8) & 0.74(0.34)\\
\noalign{\smallskip}
\hline
\end{tabular}\\
\end{center}
Fluxes are given in units of 10$^{-15}$erg~cm$^{-2}$s$^{-1}$.\\
(a) The component designation is comprised 
of a letter indicating the classification (i.e., knot $k$, hotspot $h$, lobe $l$), 
a cardinal direction plus the distance from the nucleus in arcseconds \citep{massaro11}.\\
(b) The radius column gives the size of the aperture used for photometry.\\
(c) The counts column gives the total counts in the photometric circle together with the expected background counts in parentheses; both for the 0.5 to 7 keV band.\\
{ (d) The confidence level of each detection evaluated adopting a Poisson distribution.}\\
\end{table*}

\subsection{Incidence of Intrinsic Absorption}
\label{sec:absorption}
As already investigated in Massaro et al. (2010), we performed a
photometric analysis to estimate the presence of intrinsic absorption
N$_H$ in our 3C sample.  However, our refined study is based on the
values of the hardness ratios $HR$ (see \S~\ref{sec:general})
derived from the analysis of the nuclear X-ray fluxes rather than on
the ratio between the medium and the hard X-ray fluxes as previously
performed \citep{massaro10}. Our results have
also been compared with a detailed spectral analysis that has been
performed only for the bright nuclei (see \S~\ref{sec:spectra}).

Most nuclei of radio galaxies show X-ray spectra well described by a
simple power law model with $\alpha_X$ values ranging between 0.5 and
1.5 or occasionally larger.  For those sources with little intrinsic
absorbing material (i.e. the galactic N$_{H,Gal}$ values are the major
contributors to the total absorption), the expected hardness ratio
$HR$ can be computed using the X-ray fluxes in the hard and in the
medium band.

However, we remark that if a generic source is Compton thick (i.e.,
N$_H$ > 10$^{24}$ cm$^{-2}$) or if its X-ray spectrum is inverted
(i.e. $\alpha_X$ < 0) the hardness ratios cannot provide a good
estimate of absorption. In particular, for Compton thick sources,
even if they are very rare among radio loud AGNs, the spectrum could
be dominated by a reflection component providing a low estimate of the
intrinsic absorption even if the source is heavily absorbed.  This
could also occur if the intrinsic spectrum has some features, as for
example emission lines, that only a detailed spectral fitting
procedure can reveal providing the correct estimate of N$_H$.
Thus, values of the intrinsic absorption estimated via photometry
(i.e., hardness ratios) could be different with respect to those
evaluated from spectral analysis 
{ and are thus only indicative of the presence of the absorbing material.}

On the other hand, due to the relatively short exposure times of our
snapshot survey, we often cannot recover the parameters of interest
($\alpha_X$ and N$_H$) from the spectral fits but it is possible to
derive a range of intrinsic N$_H$ column densities corresponding
to some chosen range in $\alpha_X$ by using simulated spectra.

\begin{figure}
\includegraphics[scale=0.32,origin=c,angle=-90]{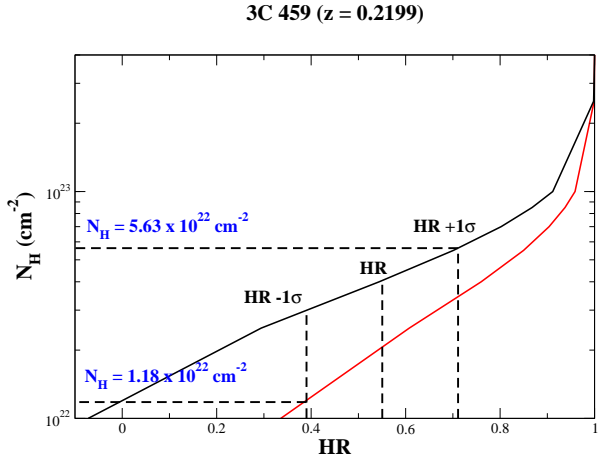}
\caption{
The relation between $HR$ and the intrinsic N$_H$ column density
resulting from the simulated spectra at redshift 0.2199 as for 3C\,459
and computed for the case of $\alpha_X$ = 0.5 (black solid line) and 1.5 (red solid line).
}
\label{fig:nh}
\end{figure}

We performed numerical simulations with {\sc xspec} deriving the values of
N$_H$ in the case of an intrinsically absorbed power-law spectrum with
different values of the spectral index $\alpha_X$ and source redshift
$z$ corresponding to different values of $HR$.  These simulated
spectra permit us to derive the relation needed to estimate the
intrinsic absorbing column density for an observed value of $HR$. 
We iterate this procedure for { two} values of $\alpha_X$ corresponding
to 0.5 and 1.5. { We note that in this photometric analysis we adopted a 
more restricted energy range of $\alpha_X$ with respect to that used in the
X-ray spectral analysis (see Section~\ref{sec:spectra})
but in agreement with our previous investigation \citep{massaro10}
and with the distribution of the spectral index of the 
low redshift 3C radio sources\citep[e.g.,][]{hardcastle09}}.
In Figure~\ref{fig:nh}, we show the N$_H$ versus $HR$
for the case of $\alpha_X$=0.5 and 1.5 for the case of 3C\,458.

Consequently, we calculated the N$_H$ estimates corresponding to the
observed $HR$, including 1$\sigma$ error, for the two values of $\alpha_X$ reported above in each source
(see Figure~\ref{fig:nh} for additional details).  
Then, we considered the maximum and the minimum values of these N$_H$ estimates to define the
range where the `real' N$_H$ value could be, 
corresponding to our estimate of the error on the intrinsic N$_H$.
{ The ranges derived for each $HR$ value are reported in Table~2 for the nuclei of AO12 sources.}

{ We repeated the entire procedure for all the sources in our
sample and for those presented in our previous work
regarding the \chn\ observations performed during Cycle 9 \citep{massaro10}.  
For the sake of consistency, we provide $HR$ and N$_H$ values in Table~3 for the nuclei of AO9 sources with less than 480 net counts (i.e those unaffected by pileup).
We found that for the Cycle 9 sources, the results on the intrinsic absorption derived with our new hardness ratio 
method are in agreement with those previously presented.}

{ Finally, we compared the results obtained from the X-ray spectral analysis with
those derived from the $HR$ study.
We found good agreement between the two methods adopted,
in fact all the values of the intrinsic absorption
derived from the X-ray spectral analysis lie within the range of N$_H$
calculated via the $HR$s.
The two cases of 3C\,79 and 3C\,234 were not compared 
because given the low numbers of counts available their spectral parameters are poorly determined.}

\subsection{Source details}
\label{sec:sources}

\noindent
\underline{{ 3C\,29}} is a nearby FR\,I radio galaxy at redshift
0.0448, showing the typical optical elliptical morphology and having
a compact nucleus in both the optical and ultraviolet.  This
source is probably an outlying member of the nearby cluster of
galaxies Abell 119, even if no significant companion appears within
$\sim$100 kpc.
The radio emission of 3C\,29 is described in Feretti et al. (1999).

Although the available radio data are not adequate to trace the S jet
back close to the nucleus, there is reasonably convincing evidence
that the inner jet is detected in the X-ray band, see
Figure~\ref{fig:3c29app}.  For a rectangular region extending to the SE
(PA=163$^{\circ}$) from just outside the nucleus for
10\arcsec, we find 15$\pm$4 net counts in the 0.5-2 keV energy range.
For the 1 - 2keV band there are 9 counts in the `on' rectangle but none
in the two adjoining background regions.  For the strongest feature
(``s2.7\arcsec\") there are 8 counts (0.5 - 2keV) within a circle of
radius=1\arcsec\ and again, none in the two adjacent
background circles.

\noindent
\underline{{ 3C\,63}} is an FR II radio galaxy that lies at redshift
0.175.  Baum et al. (1988) report on faint emission regions transverse
to the radio axis, particularly in the western region, together with
the presence of an S-shaped filament in the optical images.  Recently,
Cheung et al. (2007) have suggested that the radio morphology is that
of an ``X-shape" radio source.  
In the X-rays we detect only the nucleus.  There appears
to be diffuse X-ray emission around the nucleus up to a radius
$\approx$ 2.5\arcsec. There is also a slight X-ray excess
coincident with the brightest part of the SW radio lobe.

\noindent
\underline{{ 3C\,79}} is an FR II radio galaxy associated with a
luminous extended emission-line region \citep{fu08},  being described
as due to photoionization by the hidden active nucleus.  It has been
also classified as an HEG \citep{buttiglione09}.
{ There is no clear detection of extended emission.}

\begin{figure}
\includegraphics[scale=0.45,origin=c,angle=0]{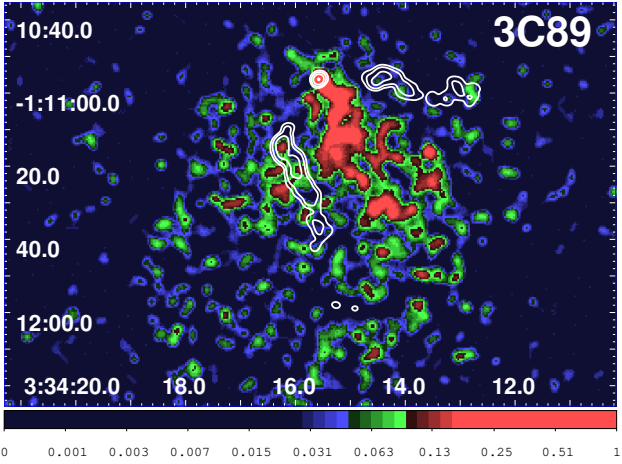}
\caption{The X-ray image of 3C89 for the energy band 0.5-7 keV.  The event file
has been smoothed with a Gaussian of FWHM=2.9\arcsec.  
Brightness units are counts per pixel. The pixel size is 0\arcsec.492.
The radio contours come from a 1.5 GHz map downloaded from the
NVAS and start at 0.5 mJy/beam, increasing by factors of four.  The
clean beam is 1.5\arcsec\ x 1.3\arcsec\ with major
axis in PA=-57$^{\circ}$.}
\label{fig:3c89a}
\end{figure}

\begin{figure}
\includegraphics[scale=0.45,origin=c,angle=0]{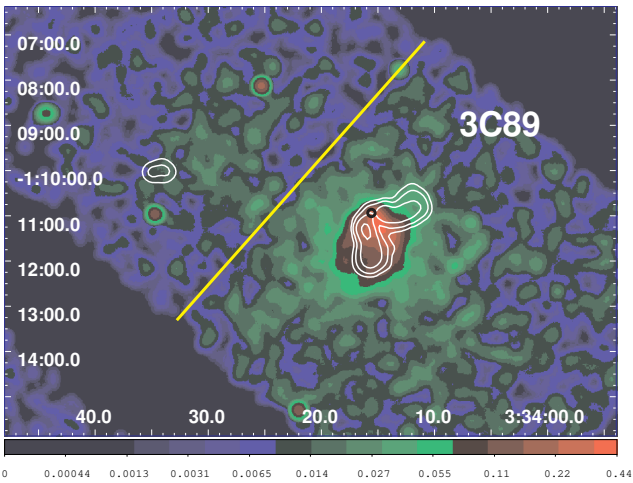}
\caption{The X-ray image of 3C89 at low resolution for the energy band 0.5-7
keV.  The event file has been regridded to a pixel size of
0.984\arcsec\ and smoothed with a Gaussian of
FWHM=10\arcsec. Brightness units are counts per pixel. The radio contours
come from a 1.5 GHz map downloaded from the NVAS and start at
5 mJy/beam, increasing by factors of four.  The clean beam is
16\arcsec\ x 13\arcsec\ with major axis in PA=
-12$^{\circ}$.   The small black circle shows the position of the host
galaxy and the yellow line follows the edge of the S3 chip.}
\label{fig:3c89b}
\end{figure}

\noindent
\underline{{ 3C\,89}} is a nearby wide angle tail radio galaxy with
an uncertain optical classification.  It is hosted in a galaxy cluster
that is clearly detected in the X-rays in our snapshot observations.
An interesting aspect is that the brighter parts of the hot gas are
between the arms of the radio source.  In particular, there is a
bright ridge of X-ray emission running back from the nucleus.  The
nuclear position is best defined by limiting the energy band to
5-7keV.  The X-ray source was detected by the ROSAT All-Sky Survey and
identified as a cluster in Appenzeller et al. (1998).  We did not
report any measure of the X-ray flux for the core since it is
contaminated by the radiation of the surrounding galaxy cluster.
{ We note that the cluster of galaxies surrounding 3C\,89 has 
the largest known X-ray cavity: $\sim$270 kpc in diameter \citep{sun12}. 
The X-ray images of 3C89 for the energy band 0.5-7 keV, with different resolution
are shown in Figure~\ref{fig:3c89a} and~\ref{fig:3c89b}}

At a redshift of 0.1386, the scale is 2.386 kpc per arcsec.  To
include all of the X-ray emission, we chose to measure gross cluster
parameters by defining a circle with radius of 2' (286 kpc).  This
circle is centered on the bright region at the center of the cluster
and just touches the edge of the chip.  We used the same size circle
for the background, shifted 4' to the SW.  For our standard band of
0.5 to 7 keV, we find 1858$\pm$58 net counts.  We measured fluxmaps
and find a total of 1.50$\times$10$^{-12}$ erg~cm$^{-2}$~s$^{-1}$,
corresponding to a luminosity of 7.3$\times$10$^{43}$~erg~s$^{-1}$.
The uncertainties in these values is of order 7\%.

Using the same regions, we performed a standard spectral analysis
using Sherpa with the absorption frozen at the galactic value.  Since
we noted that the flux distribution was weighted towards the higher
energies (more than half of the total flux is above 2 keV), we tried
fits for both an absorbed power law and an absorbed APEC model.  Both
fits gave reduced $\chi^2$ values less than 1.  The spectral index for
the power law fit is 0.62$\pm$0.08 and the APEC model (for a frozen
solar abundance of unity) gives kt=5.7$\pm$1.6 keV.  Approximately 1/5
of the total luminosity comes from the bright central region between
the radio tails.

\noindent
\underline{{ 3C\,93.1}} is a CSS radio source that shows an extended
optical narrow-line emitting region surrounding the core for about 3.8
kpc, primarily toward the north-east side \citep{tremblay09}.  We
clearly detected the X-ray core but we did not find any evidence of
extended structures.

\noindent
\underline{{ 3C\,130}} is a classical FRI radio galaxy at redshift
0.1090.  The optical morphology is clearly elliptical without any
peculiarity.  The core is detected in the X-rays and it also shows
significant extended X-ray emission around its central region.

\noindent
\underline{{ 3C\,166}}: is an FR\,II radio galaxy optically
classified as LEG \citep{buttiglione09}.  It shows an unusual radio
structure featuring two lobes with very different morphologies
\citep{spangler82}.  The source has a bright unresolved infrared
nucleus \citep{floyd08}.  We did not detect any extended emission in
the X-ray.

\noindent
\underline{{ 3C\,180}} is a classical FR\,II radio source optically
classified as HEG \citep{buttiglione09}. 
It is hosted in a giant elliptical galaxy \citep{madrid06} and according to McCarthy et
al. (1995) it is also a member of a cluster of galaxies.  
{ We do not detect 
any clear signatures of X-ray emission arising from the surrounding galaxy cluster.}

\noindent
\underline{{ 3C\,196.1}} lies within a cluster of
galaxies catalogued in the ROSAT bright source catalog and by Kocevski
et al. (2007) in their search for clusters in the zone of avoidance.
With an exposure of 8ks, there are 3700 net counts detected in a
circle of radius 90\arcsec.  The global temperature estimate
for the entire cluster is 4.2 $\pm$ 0.2 keV.  Restricting the spectral
analysis to smaller circles, the temperature drops: 3.4 $\pm$ 0.2 for
r=20\arcsec; 3.0 $\pm$ 0.3 for r=7\arcsec, and 1.0
keV for r=1.5\arcsec, the last value centered on the
brightest part of the inner structure.  With a redshift of 0.198, the
luminosity distance is 936 Mpc and the angular scale is 3.16 kpc per
arcsec.  The luminosity in the band 0.5 to 7 keV is 3$\times$~10$^{44}$
erg s$^{-1}$.  Some further aspects of this cluster have been discussed
by Harris et al. (2011).

\begin{figure}
\includegraphics[scale=0.45,origin=c,angle=0]{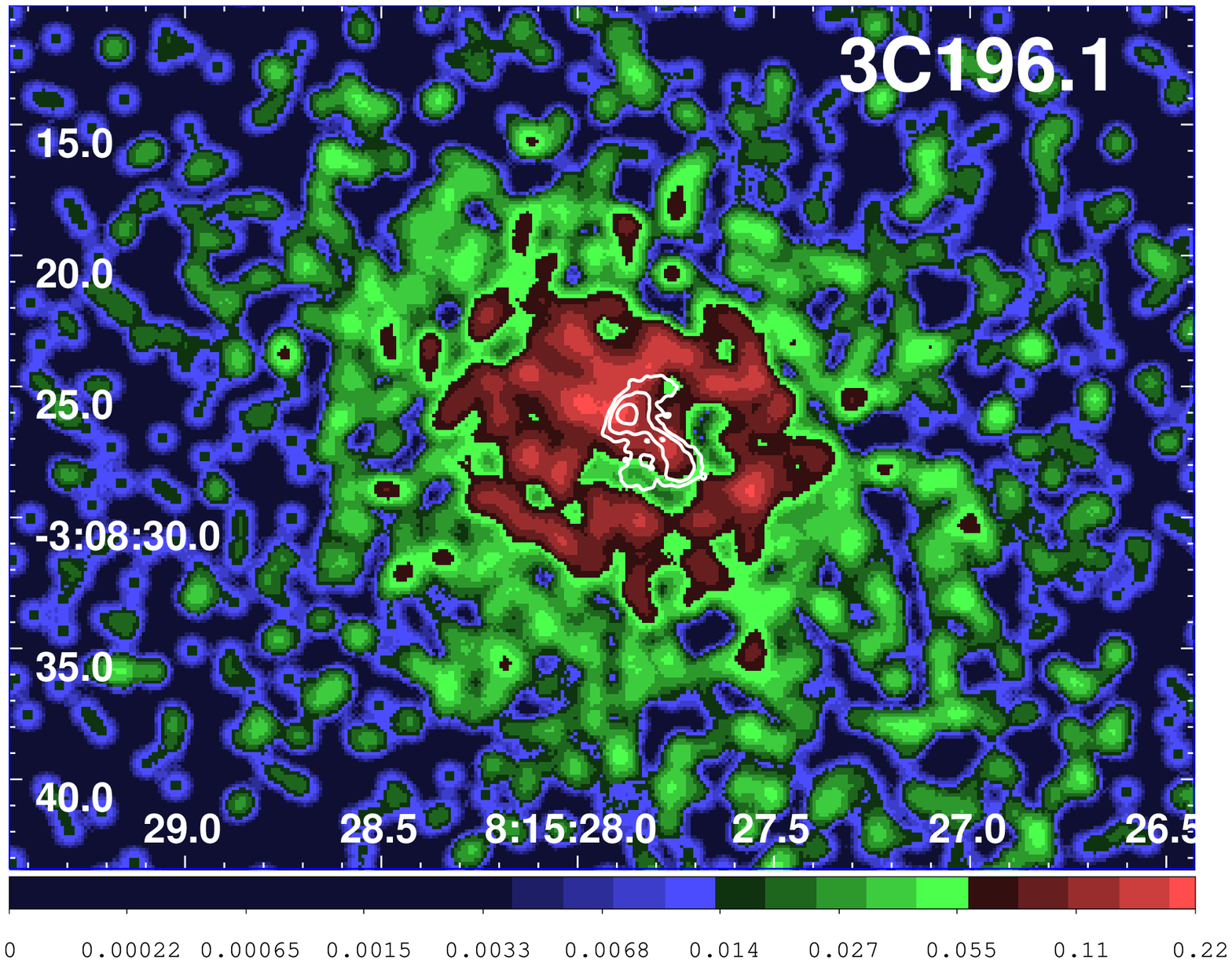}
\caption{The X-ray image of the inner part of the cluster emission
from 3C196.1 for the energy band 0.5-7 keV.  The event file has been
regridded to a pixel size of 0.123\arcsec\ and smoothed with
a Gaussian of FWHM=1.0\arcsec. Brightness units are counts
per pixel.  The radio contours come from an 8.4 GHz map kindly
provided by C. C. Cheung, and start at 0.5 mJy/beam, increasing by
factors of four.  The clean beam is 0.3\arcsec.}
\label{fig:3c196.1a}
\end{figure}

\begin{figure}
\includegraphics[scale=0.45,origin=c,angle=0]{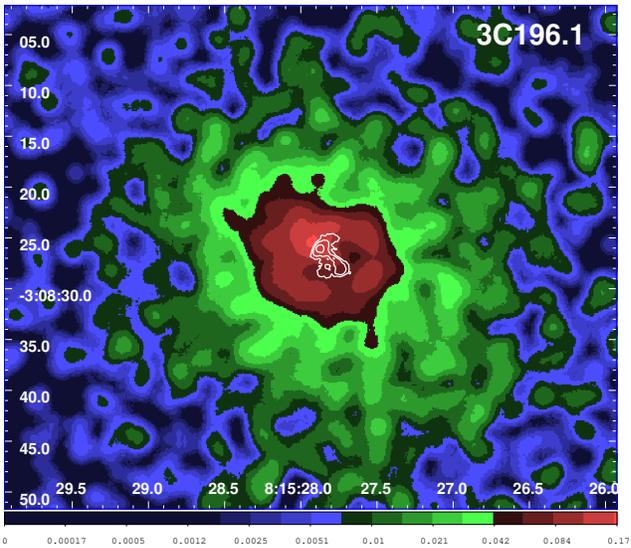}
\caption{An low resolution X-ray image of 3C196.1 for the energy band
0.5-7 keV.  The event file has been regridded to a pixel size of
0.123\arcsec\ and smoothed with a Gaussian of
FWHM=2.2\arcsec. Brightness units are counts per pixel.  The
radio contours come from an 8.4 GHz map kindly provided by
C. C. Cheung, and start at 0.5 mJy/beam, increasing by factors of
four.  The clean beam is 0.3\arcsec.}
\label{fig:3c196.1b}
\end{figure}

The radio source has been classified as an FRII and is associated with
the dominant galaxy of the cluster.  The radio power and optical
magnitude of 3C\,196.1 place it well above the dividing line between FRI
and FRII of Owen \& Laing (1989).  However, its morphology is
strikingly different from the prototype FRII, Cygnus A, which also
resides in a cluster.  Unlike Cygnus A with a total size of 120 kpc (from
hotspot to hotspot), the extent of 3C196.1 is only 12 kpc.  Thus it
probably is wholly within the cD galaxy, although we cannot be sure of
this since the irregular morphology precludes even a reasonable guess
at projection effects.  The physical size of 3C196.1 could be much
larger than its projected size.  The best description of this source
with the available 0.3'' resolution is a 'HYMORS', a so called 'HYbrid
Morphology Radio Source' \citep{gopal00,kharb10}.  In the case of 3C196.1, the
SW side is jet like (i.e. FRI) whereas the NE side appears to be a
classical FRII lobe with a brightness enhancement towards the
edge. Perhaps however, this feature is not a true hotspot, but rather
the location where the NE jet impacts the higher density ICM, as
indicated by the peak X-ray brightness.  In Figure~\ref{fig:3c196.1a}
\&~\ref{fig:3c196.1b}, there appears to be a
`ghost cavity' wrapping around the SW jet/lobe.  The steep (radio)
spectrum 'S lobe' impinges on part of this cavity.

Since the Galactic latitude of 3C196.1 is 17$^\circ$, the
optical attributes of the cluster are not well determined.  The host
galaxy is a cD, the dominant galaxy of a group of 14 others that lie
within about 350 kpc from its core \citep{baum88}. Madrid et
al. (2006) show that its near-infrared image is elliptical, presenting
elongated structure northeast to southwest, which is the same
direction as the radio emission. The same morphology is seen in the
optical \citep{dekoff96,baum88}. [OIII] emission lines are observed in
two localized regions cospatial with the elongated bright core
\citep{tremblay09}.
 
\noindent
\underline{{ 3C\,198}} is classified as FR\,II on the basis of its radio luminosity at 178 MHz,
as reported in Spinrad et al. (1985). { With the currently available radio data we have been unable to locate
a radio nucleus. The data in Table 2 come from measuring a weak X-ray source at the NED position of the optical galaxy.}

\noindent
\underline{{ 3C\,223}} is a classical FR\,II - HEG radio galaxy
whereas the near-infrared image shows an asymmetric galaxy, slightly
elongated from the northeast to southwest side \citep{madrid06}.  We
detected only the X-ray core in our \chn\ snapshot survey.

\noindent
\underline{{ 3C\,234}} is an FR\,II radio galaxy optically
classified as HEG.  Its optical images \citep{tremblay09} show an
elongated emission line region that includes a tidal arm
\citep{carleton84} that appears to be roughly parallel with that of
the radio jet axis on large scales \citep[see also][for more
details]{privon08}.   The SW hotspot is clearly detected in
the X-ray band.
{ The X-ray spectral analysis of the nuclear emission in 3C\,234 has been presented by
Piconcelli et al. (2008). They found a thermal dominated X-ray spectrum, rich of emission lines,
confirming the presence of a hidden quasar in this source and
claiming that the origin of the X-ray emission in
radio-loud AGNs with high-excitation optical lines 
does not always arise from jet non-thermal radiation.}
 
\noindent
\underline{{ 3C\,258}} This galaxy is a well-known compact
steep-spectrum (CSS) radio galaxy with an uncertain redshift estimate
\citep{floyd08}.  
 
\noindent
\underline{{ 3C\,284}} is a classical FR\,II - HEG radio source,
hosted by a disturbed elliptical galaxy at redshift 0.2394 with a
southeast tidal tail toward its most prominent companion
\citep{floyd08}.  
Only the X-ray core is detected in our \chn\ observation.
 
\noindent
\underline{{ 3C\,314.1}} is an FR\,I - LEG radio galaxy hosted in an
elliptical galaxy that shows an infrared elongated structure from the
east-northeast side to west-southwest \citep{madrid06}.  

\noindent
\underline{{ 3C\,319}} is an FR\,II radio galaxy, the only
source in our current sample for which we did not detect X-ray
emission from the nucleus.  The NE hotspot is detected with 4 counts.
The galaxy is also very faint in the HST optical images with an extended emission line region dominated by the
[O III] line that is more compact than the H$\alpha$+[N II] one
\citep{tremblay09}.
 
\noindent
\underline{{ 3C\,357}} is an FR\,II - LEG radio galaxy at
redshift 0.1662 with a projected angular size corresponding to
$\sim$250 kpc \citep{fanti97,harvanek98}.  It shows an optical
emission line region of a few arcsec scale extending from the nucleus
along the radio axis (McCarthy et al. 1995). Both de Koff et
al. (1996) and Capetti et al. (2000) reported on filamentary dust
lanes in the southwestern region.  The host galaxy is a classical
giant elliptical with no remarkable features \citep{floyd08}.  In our
\chn\ snapshot observation we clearly detected the core emission but
no extended radiation around the core or associated with the hotspots
has been observed.

\noindent
\underline{{ 3C\,379.1}} is an FR\,II - HEG radio galaxy hosted in
an elongated elliptical galaxy as seen in the near IR HST images
\citep{floyd08}. 

\noindent
\underline{{ 3C\,402}} There are two bright galaxies separated by
$\approx$2$^{\prime}$ in declination and each appears to be a radio
galaxy.  We suspect that the large scale radio structures are
superposed.  Both nuclei are detected in radio and X-rays and both are
somewhat extended.  With the radio maps available to us, our
registration of the X-ray map is less secure than usual. 
{ This is because of the difficulty of determining
the location of the peak brightness for an extended feature and
because the radio maps have a large beam size.}
The X-ray nucleus of 402N has a 2\arcsec\ long extension to the SW,
which could be a jet.

\noindent
\underline{{ 3C\,403.1}} is a double FR\,II radio source with an LEG
optical classification \citep{buttiglione09}.  
 
\noindent
\underline{{ 3C\,410}} is an FR\,II broad line radio galaxy
\citep{buttiglione09} with a redshift of 0.2485.  The X-ray nucleus
has to be shifted by 1\arcsec\ in declination in order to
align the central component of the radio triple with the X-ray. In our
experience in registering well over a hundred \chn\ observations,
the usual shift required to align radio and X-ray nuclei is between
0.2\arcsec\ and 0.3\arcsec, and almost never larger
than 0.4\arcsec\ (see also Section~\ref{sec:obs}).
{ However, since the absolute pointing determined by the Chandra star
trackers is occasionally off by $\approx 1^{\prime\prime}$ (T. Aldcroft,
personal communication), we believe the registration performed is
correct.}

\noindent
\underline{{ 3C\,424}} is a nearby FR\,I - LEG that lies at redshift
0.127 in a dense environment, with numerous companions clearly
detected in the optical HST image \citep{dekoff96}. { The radio source
apparently is close to the edge of a galaxy cluster \citep{dekoff96}, }
however cluster emission is not detected in the X-ray band.  
The X-ray nucleus has an extension 0.6\arcsec\ to the NW which 
might be the inner segment of a jet.

\noindent
\underline{{ 3C\,430}} is an FR\,II - LEG radio galaxy
hosted in an elongated elliptical galaxy at redshift 0.0541, with a
dust lane around the nucleus that could be responsible for
absorption in the X-ray band as suggested by the lack of X-ray counts
in the soft band.
 
\noindent
\underline{{ 3C\,436}} is a typical FR\,II - HEG hosted in { an}
elliptical galaxy elongated in the same direction as its radio
structure \citep{madrid06}.  For the nucleus, 25 of the 41 total
counts lie between 4 and 6 keV.  We detect what might be described as
the 'primary' radio hotspot close to the tip of the S lobe.

\noindent
\underline{{ 3C\,456}} is an FR\,II - HEG radio galaxy at redshift 0.233.
In the X-rays, we detected only the nucleus which appears to be highly absorbed 
(see also \S~\ref{sec:spectra} for additional details).

\noindent
\underline{{ 3C\,458}} We detect the NE hotspot with 4 counts { in this FR\,II - HEG radio galaxy}.

\noindent
\underline{{ 3C\,459}} is an FR\,II broad line radio galaxy.  The host galaxy is
dominated by a young stellar population \citep{tadhunter02} and its IR
structure has been suggested to be the result of a recent merger
\citep{floyd08}. The radio source is small and very asymmetric
\citep{morganti99}.  The high resolution radio map \citep{morganti93}
reveals a weak western component; see Figure~\ref{fig:3c459app}.  The
extended X-ray emission around the nucleus is mostly on the E side,
coincident with the E radio lobe.

\section{Summary}
\label{sec:summary}
We have presented our X-ray analyses of the second half of the 3C low
redshift sample (i.e., $z<$0.3) observed by \chn\ during Cycle
12. Since we waived proprietary rights, X-ray data for all
extragalactic 3C sources with z$<$0.3 are now available to the
community for statistical analyses based on a complete, unbiased
sample.  In addition, we have found several sources worthy of more
detailed study such as 3C\,89 for which follow up \chn\ observations
totaling 68ks with ACIS-I have been already performed in Cycle 13
\citep{sun12}.

We have constructed fluxmaps for all the X-ray observations and given
photometric results for the nuclei and other radio structures (i.e.,
jet knots, hotspots, lobes).  For the stronger nuclei, we have
employed the usual { X-ray} spectral analysis, and compared the column
densities of intrinsic absorption to those obtained from the hardness
ratio analysis (see \S~\ref{sec:absorption}).  As expected, X-ray
emission was detected for all the nuclei except for 3C\,319, a small
FR II radio galaxy. A sizable fraction ($\sim$1/3) of our 3C sources
show evidence for significant intrinsic absorption (see
\S~\ref{sec:spectra} and \S~\ref{sec:absorption} for more details).

Amongst our 3C \chn\ observations, we detected two compact steep
spectrum radio sources: 3C\,93.1 and 3C\,258, one wide angle tail radio
galaxy, 3C\,89, hosted in the cluster of galaxies with the largest
known X-ray cavity \citep{sun12} and the X-ray emission of the galaxy
cluster surrounding 3C\,196.1.  We also detected X-ray emission from three
radio hotspots and, in the case of 3C\,459, emission coincident with
the eastern radio lobe (see \S~\ref{sec:results}).  Finally, we
found X-ray emission cospatial with 2 radio jets: 3C\,29 and 3C\,402.

\acknowledgments 
We thank the anonymous referee for useful comments that led to improvements in the paper.
We wish to honor the memory of our great friend and colleague David Axon, who has been the steadfast inspiration and participant 
in this and many other key papers that through many years of dedicated efforts have led to significant breakthroughs 
and greater understanding of the physics of active galaxies. He will be greatly missed by all of us.
We are grateful to M. Hardcastle and C. C. Cheung for providing
several radio maps of the 3C sources.  We also thank C. C. Cheung
and S. Bianchi for helpful discussions.  This research has made
use of NASA's Astrophysics Data System; SAOImage DS9, developed by the
Smithsonian Astrophysical Observatory; and the NASA/IPAC Extragalactic
Database (NED) which is operated by the Jet Propulsion Laboratory,
California Institute of Technology, under contract with the National
Aeronautics and Space Administration.  Several radio maps were
downloaded from the NVAS (NRAO VLA Archive Survey)
and from the DRAGN webpage\footnote{http://www.jb.man.ac.uk/atlas/}.  
The National Radio Astronomy Observatory is operated by Associated Universities,
Inc., under contract with the National Science Foundation.  A few
radio maps have been obtained from the Merlin archive.  The work at
SAO is supported by NASA-GRANT GO8-9114A and the work at RIT was
supported by \chn\ grant GO8-9114C.  F. Massaro acknowledges the
Foundation BLANCEFLOR Boncompagni-Ludovisi, n'ee Bildt for the grant
awarded him in 2009 and in 2010.
This work is supported in part by the Radcliffe Institute for Advanced Study at Harvard University.

{ Facilities:} \facility{VLA}, \facility{CXO (ACIS)}

\appendix
\section{Images of the sources}
\label{sec:appendix}
Although for many of our sources the X-ray data are comprised of
rather few counts, we show here the radio morphology via contour
diagrams which are superposed on X-ray event files that have been
smoothed with a Gaussian.  The full width half maximum (FWHM) of the
Gaussian smoothing function is given in the figure captions.  When there is
sufficient S/N (signal to noise ratio) of the X-ray image to provide spatial information, we
have added contours (cyan or white)
which are normally separated by factors of two.  Most of the overlaid
radio contours increase by factors of four.
The X-ray event files shown are in units of counts/pixel in the 0.5-7 keV energy range.
The primary reason figures appear so different from each other is the
wide range in angular size of the radio sources.  

\begin{figure}
\includegraphics[keepaspectratio=true,scale=0.90]{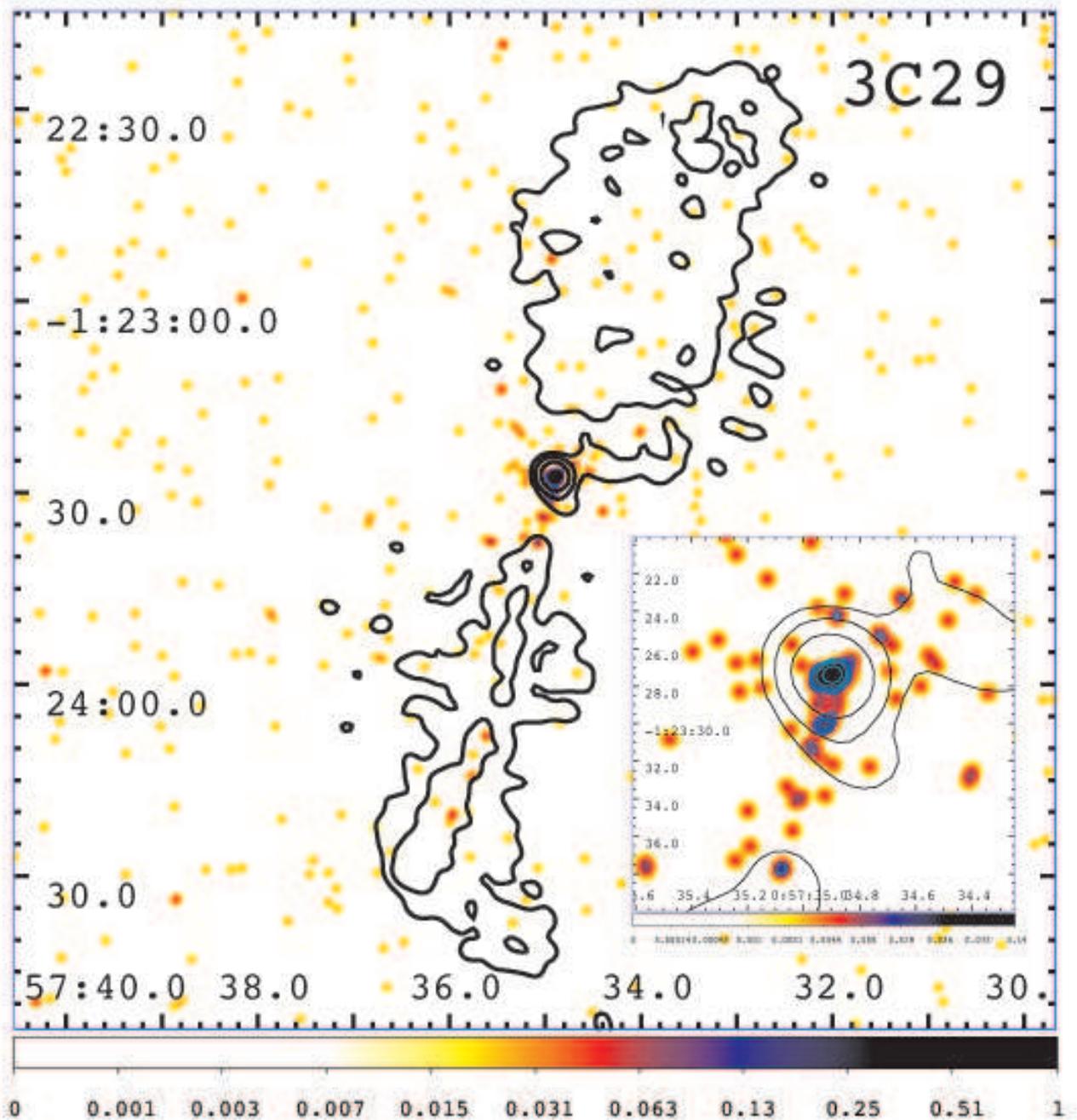}
\caption{The X-ray image of 3C29 for the energy band 0.5-7 keV.  The event file
has been regridded to a pixel size of 0.246\arcsec\ and
smoothed with a Gaussian of FWHM=1.4\arcsec.  For the
insert, the pixel size is 0.0615 and the FWHM of the smoothing
function is 0.8\arcsec.  X-ray contours (white or cyan)
start at 0.01 counts/pix and increase by factors of two.  The radio
contours (black) come from an 8.4 GHz map downloaded from the NVAS and
start at 0.9 mJy/beam, increasing by factors of four.  The clean beam
is 3.0\arcsec\ x 2.7\arcsec\ with major axis in
PA=43$^{\circ}$.}
\label{fig:3c29app}
\end{figure}

\clearpage
\begin{figure}
\includegraphics[keepaspectratio=true,scale=0.90]{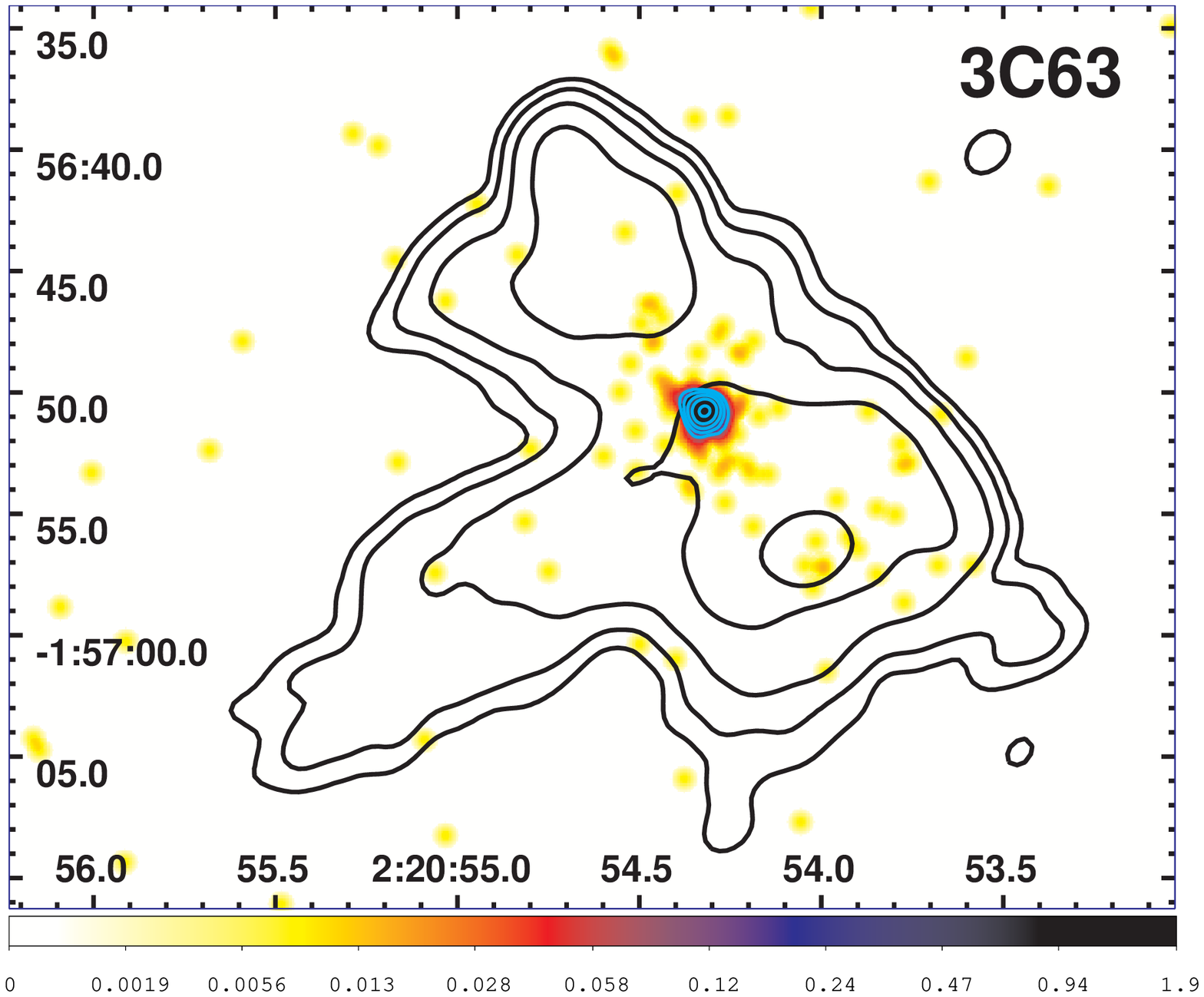}
\caption{The X-ray image of 3C63 for the energy band 0.5-7 keV.  The event file
has been regridded to a pixel size of 0.0615\arcsec\ and
smoothed with a Gaussian of FWHM=0.65\arcsec.  X-ray
contours (white or cyan) start at 0.1 counts/pix and increase by
factors of two.  The radio contours (black) come from a 1.4 GHz map
downloaded from the NVAS and start at 0.4 mJy/beam, increasing by
factors of four.  The clean beam is
1.9\arcsec\ x 1.4\arcsec.}
\label{fig:3c63app}
\end{figure}

\clearpage
\begin{figure}
\includegraphics[keepaspectratio=true,scale=0.90]{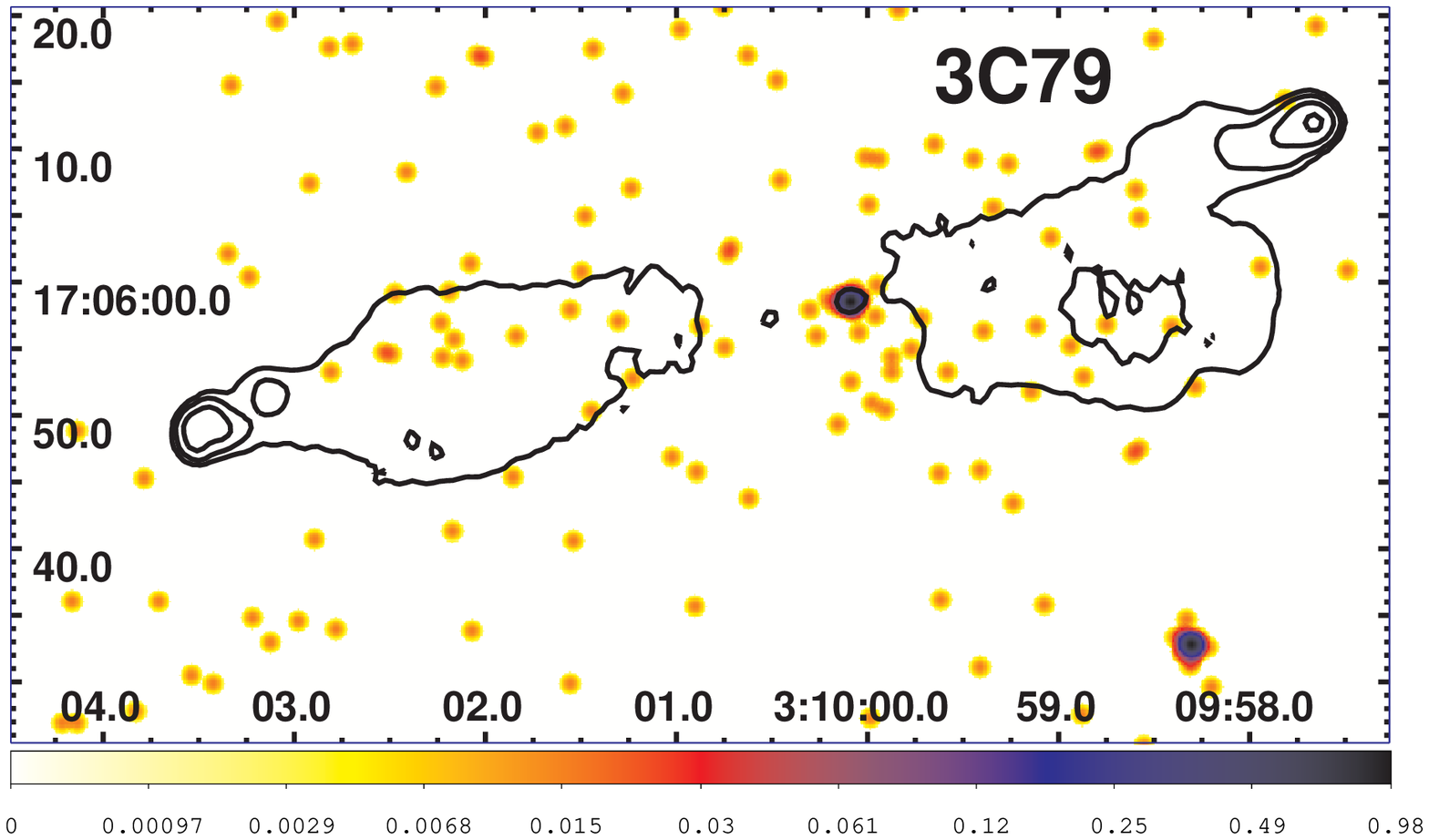}
\caption{The X-ray image of 3C79 for the energy band 0.5-7 keV.  The event file
has been regridded to a pixel size of 0.123\arcsec\ and
smoothed with a Gaussian of FWHM=1.0\arcsec.  The radio
contours (black) come from a 1.4 GHz map downloaded from the DRAGN
website, and start at 4 mJy/beam, increasing by factors of four.  The
clean beam is 1.5\arcsec.}
\label{fig:3c79app}
\end{figure}

\clearpage
\begin{figure}
\includegraphics[keepaspectratio=true,scale=0.90]{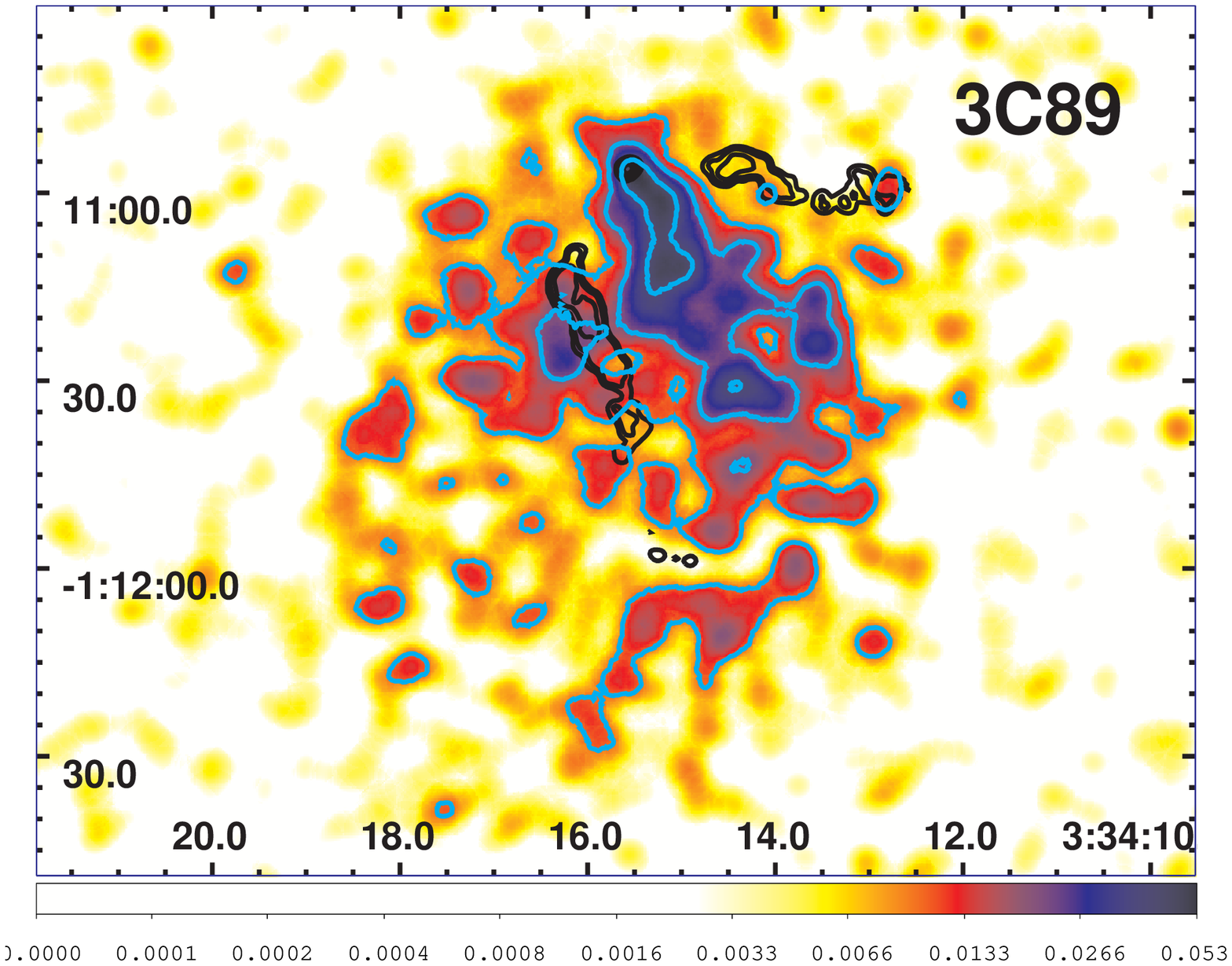}
\caption{The X-ray image of 3C89 for the energy band 0.5-7 keV.  The event file
has been regridded to a pixel size of 0.246\arcsec\ and
smoothed with a Gaussian of FWHM=5.5\arcsec.  X-ray contours
(white or cyan) start at 0.01 counts/pix and increase by factors of
two.  The radio contours (black) come from a 1.5 GHz map downloaded
from the NVAS and start at 0.5 mJy/beam, increasing by factors of
four.  The clean beam is 1.5\arcsec\ x 1.3\arcsec
FWHM with major axis at PA=-57$^{\circ}$.}
\label{fig:3c89app}
\end{figure}

\clearpage
\begin{figure}
\includegraphics[keepaspectratio=true,scale=0.90]{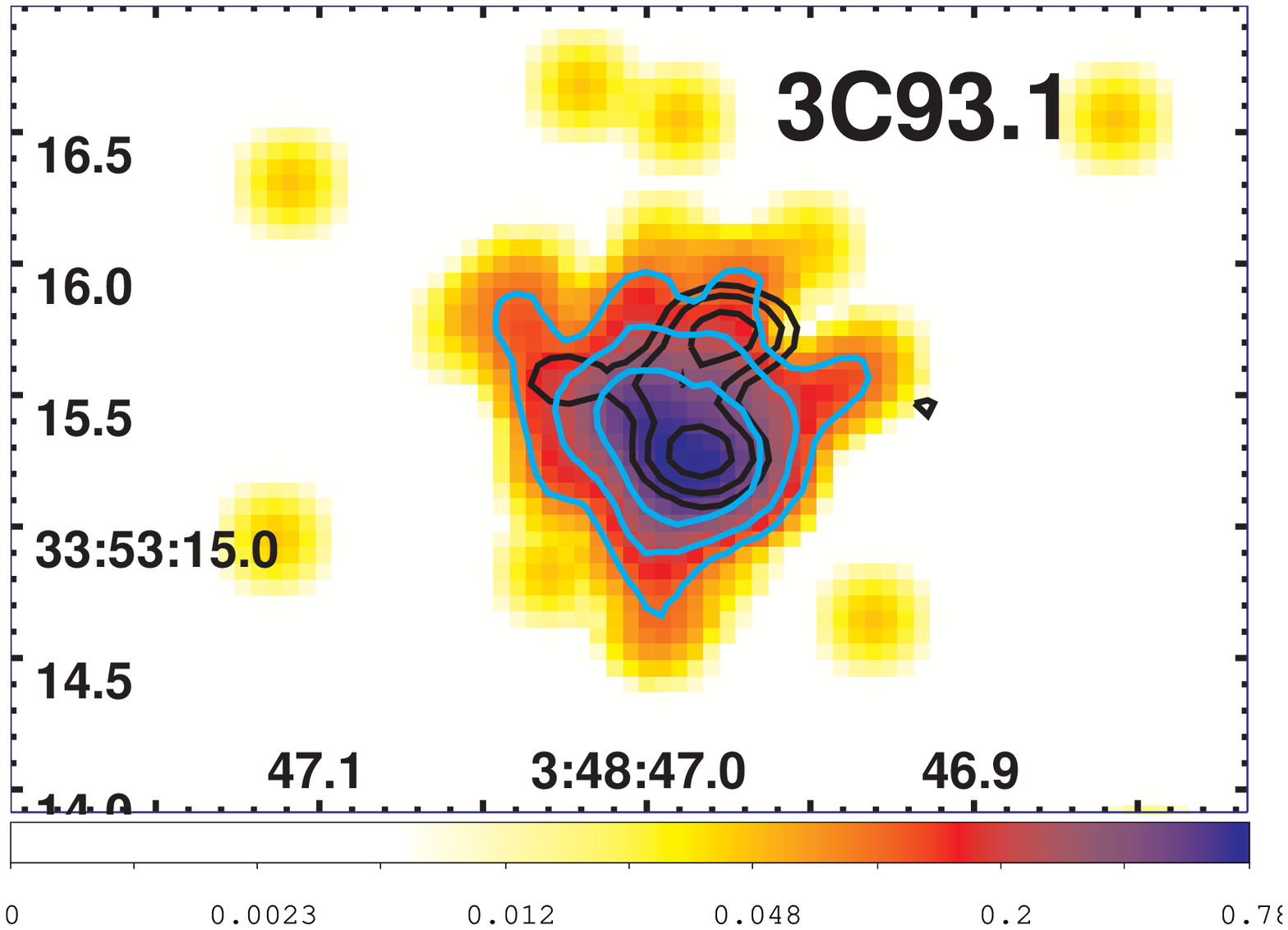}
\caption{The X-ray image of 3C93.1 for the energy band 0.5-7 keV.  The event
file has been regridded to a pixel size of 0.0615\arcsec\ and
smoothed with a Gaussian of FWHM=0.3\arcsec.  X-ray contours
(white or cyan) start at 0.1 counts/pix and increase by factors of
two.  The radio contours (black) come from a 8.4 GHz map downloaded
from the NVAS and start at 10 mJy/beam, increasing by factors of two.
The clean beam is 0.3\arcsec\ x 0.2\arcsec\ with major
axis at PA=90$^{\circ}$.}
\label{fig:3c93.1app}
\end{figure}

\clearpage
\begin{figure}
\includegraphics[keepaspectratio=true,scale=0.90]{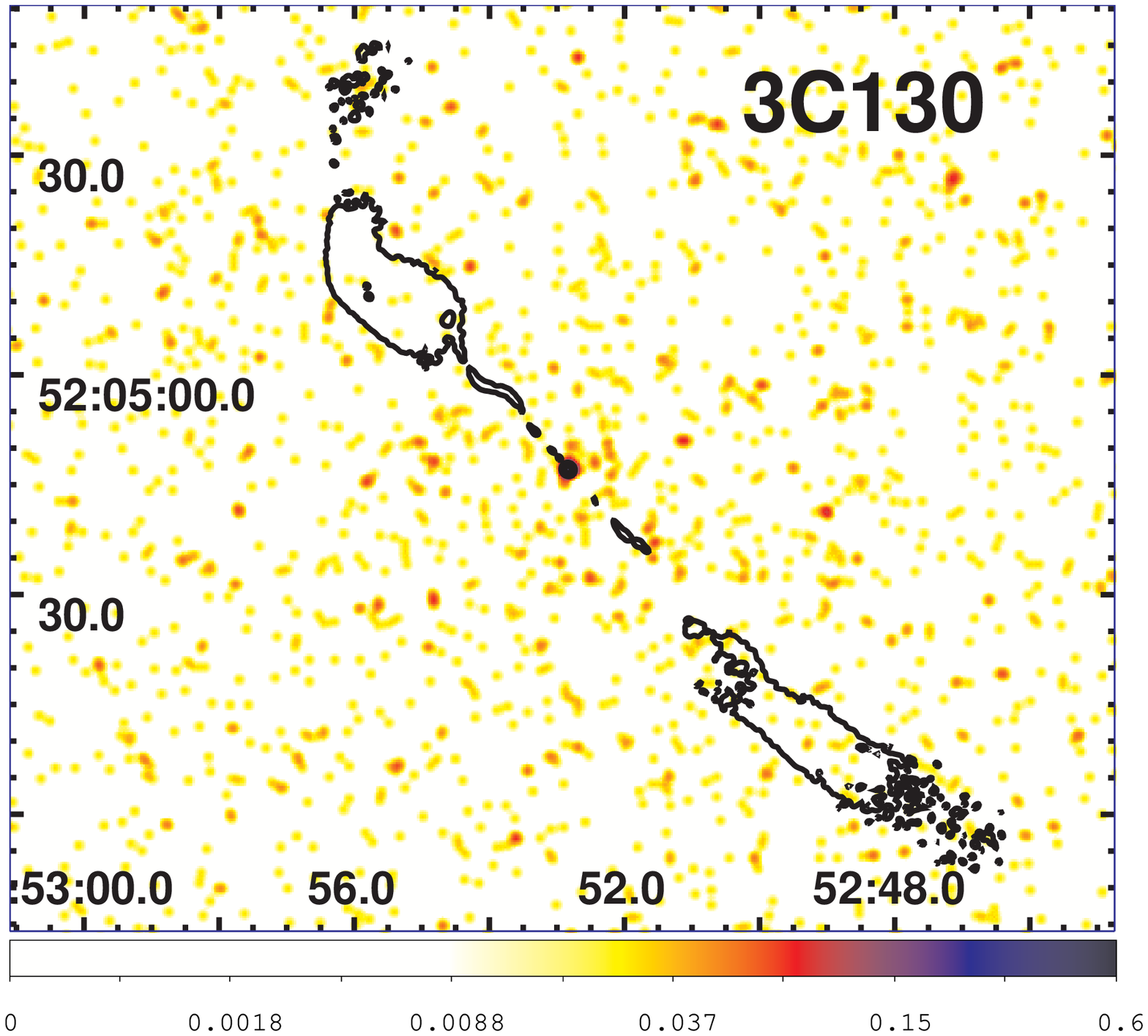}
\caption{The X-ray image of 3C130 for the energy band 0.5-7 keV.  The event
file has been regridded to a pixel size of 0.246\arcsec\ and
smoothed with a Gaussian of FWHM=1.4\arcsec.  The radio
contours (black) come from a 8.4 GHz map kindly supplied by
M. Hardcastle and start at 0.1 mJy/beam, increasing by factors of
four.  The clean beam is 0.6\arcsec.}
\label{fig:3c130app}
\end{figure}

\clearpage
\begin{figure}
\includegraphics[keepaspectratio=true,scale=0.90]{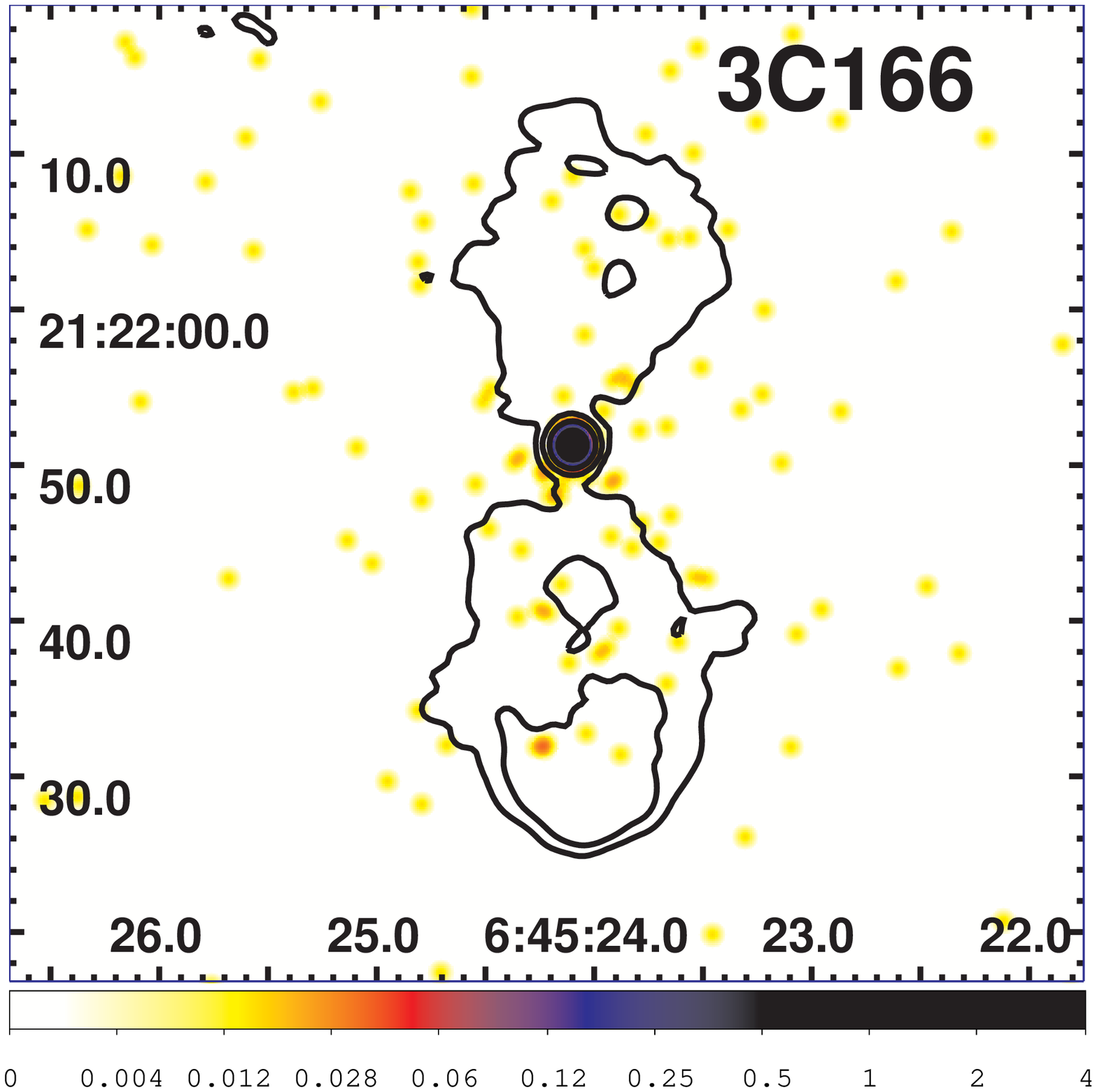}
\caption{The X-ray image of 3C166 for the energy band 0.5-7 keV.  The event
file has been regridded to a pixel size of 0.123\arcsec\ and
smoothed with a Gaussian of FWHM=1.0\arcsec.  The radio
contours (black) come from a 1.4 GHz map downloaded from NED,
and start at 2 mJy/beam, increasing by factors of four.  The clean
beam is 1.3\arcsec.}
\label{fig:3c166app}
\end{figure}

\clearpage
\begin{figure}
\includegraphics[keepaspectratio=true,scale=0.90]{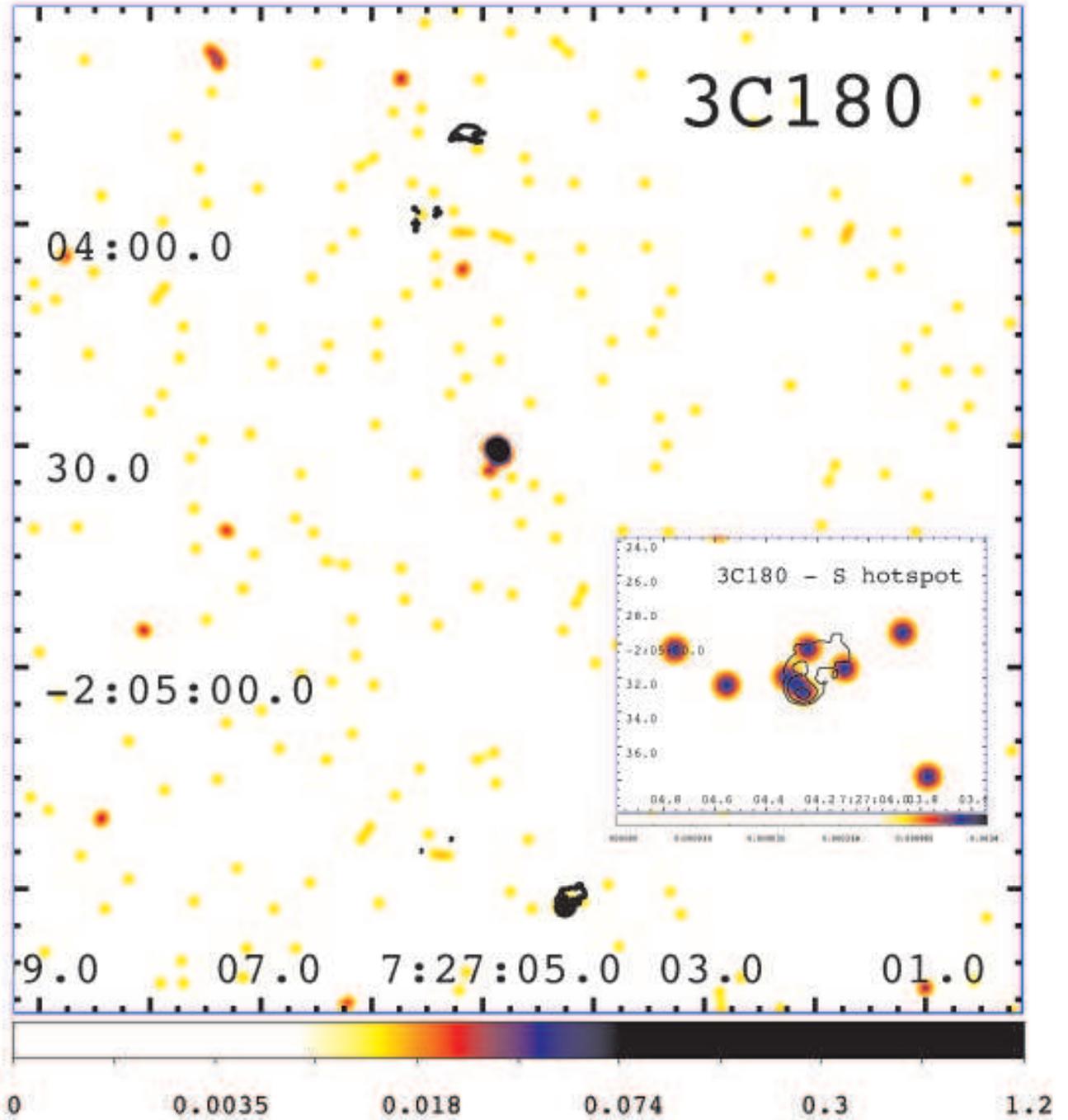}
\caption{The X-ray image of 3C180 for the energy band 0.5-7 keV.  The event
file has been regridded to a pixel size of 0.246\arcsec\ and
smoothed with a Gaussian of FWHM=2\arcsec.  The radio
contours (black) come from a 8.4 GHz map kindly supplied by
C. C. Cheung and start at 0.5 mJy/beam, increasing by factors of four.
The clean beam is 0.36\arcsec.  For the insert of the S
hotspot, the pixel size is 0.06\arcsec\ and the smoothing
function has a FWHM=1.4\arcsec.}
\label{fig:3c180app}
\end{figure}

\clearpage
\begin{figure}
\includegraphics[keepaspectratio=true,scale=0.90]{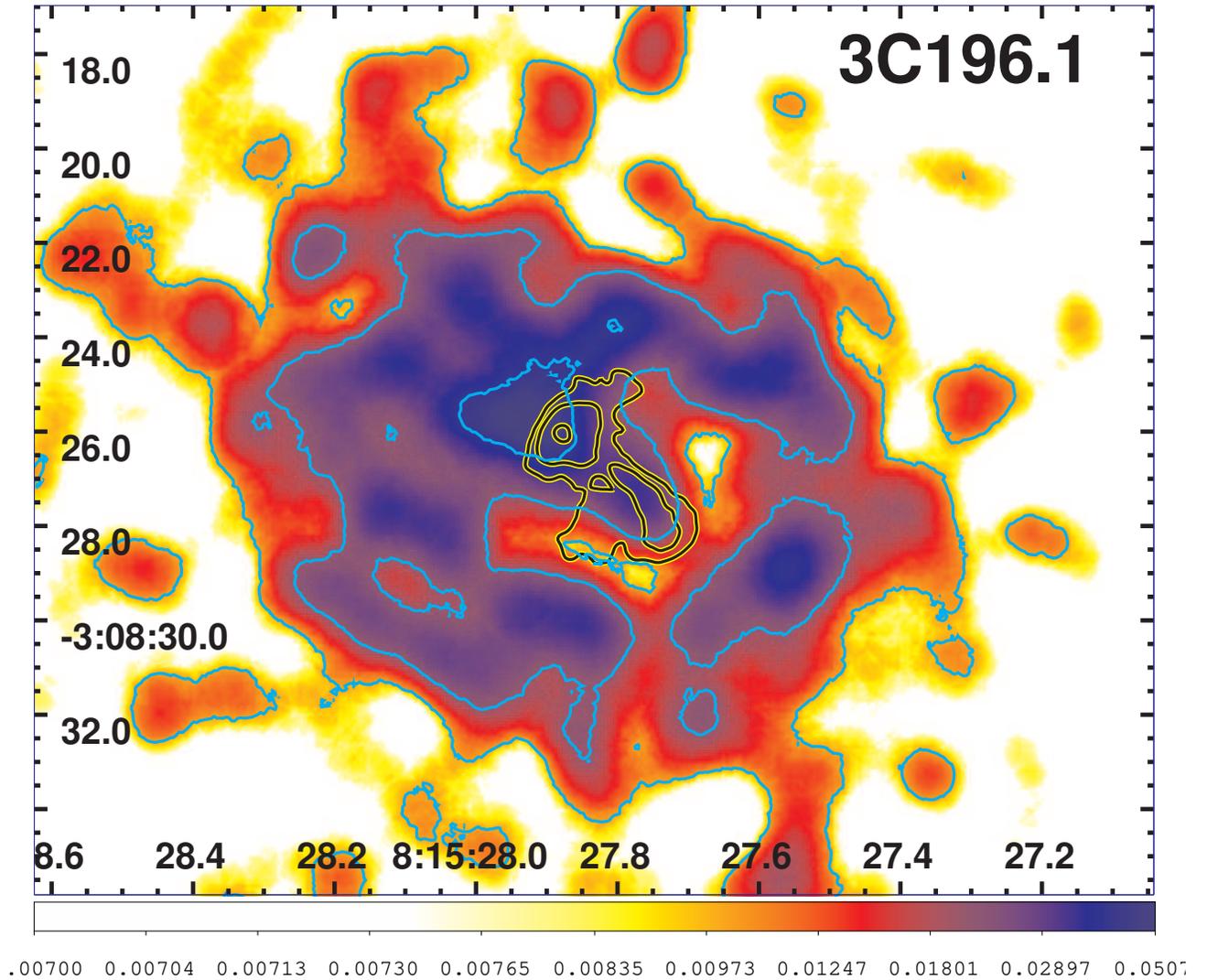}
\caption{The X-ray image of 3C196.1 for the energy band 0.5-7 keV.  The pixel
size is 0.492\arcsec\ and the map has been smoothed with a
Gaussian of FWHM=11\arcsec.  X-ray contours (white or cyan)
start at 0.01 counts/pix and increase by factors of two.  The radio
contours (black, outlined in yellow) come from an 8.4 GHz map kindly
supplied by C. C. Cheung and the contours start at 0.75 mJy/beam,
increasing by factors of four.  The clean beam is
0.3\arcsec.}
\label{fig:3c196.1app}
\end{figure}

\clearpage
\begin{figure}
\includegraphics[keepaspectratio=true,scale=0.90]{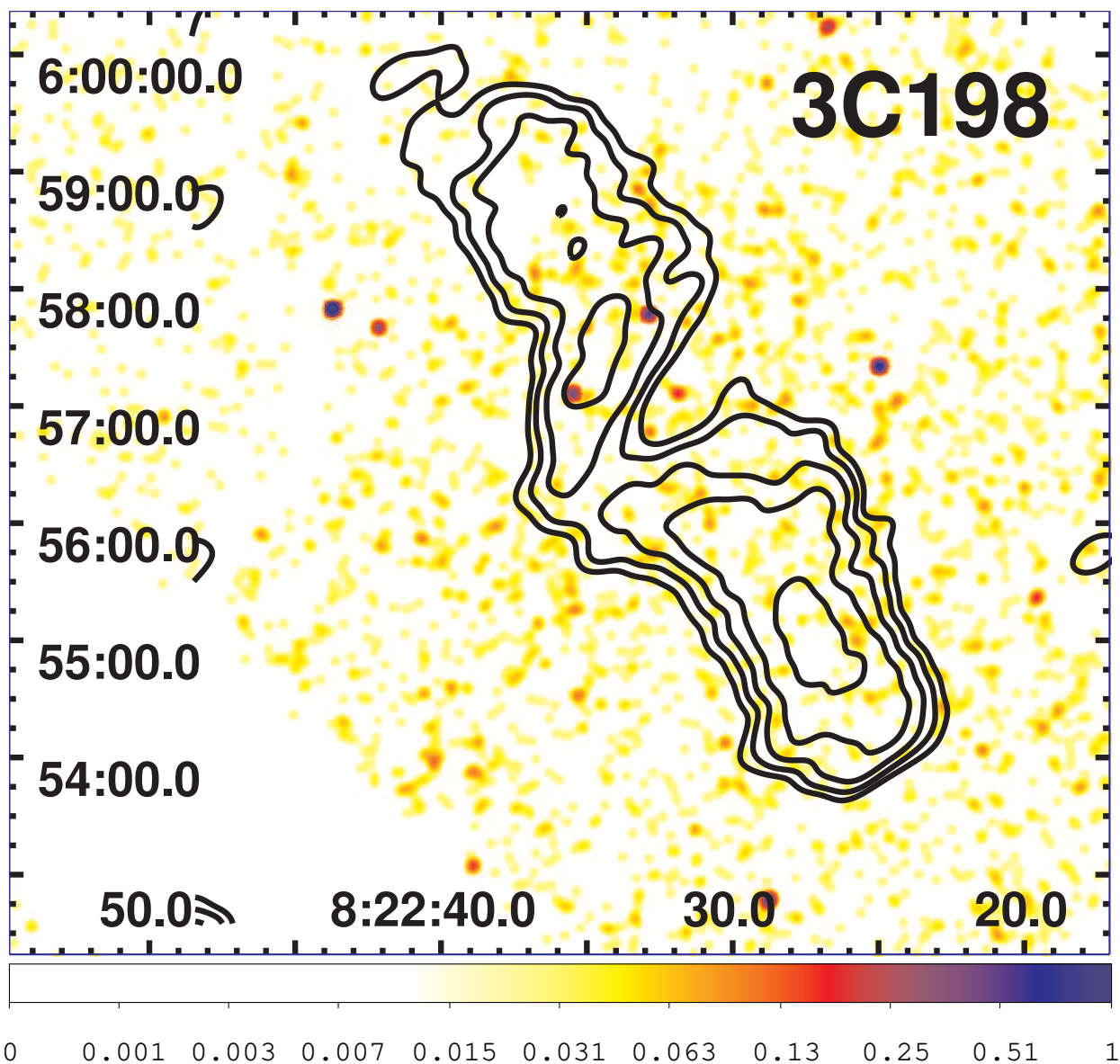}
\caption{The X-ray image of 3C198 for the energy band 0.5-7 keV.  The event
file has been blocked by a factor of two so the pixel size is
0.984\arcsec.  The map has been smoothed with a Gaussian of
FWHM=11.8\arcsec.  The radio contours (black) come from a
4.9 GHz map produced from the NRAO archives, and start at 0.5
mJy/beam, increasing by factors of two.  The clean beam is
25\arcsec\ x 14\arcsec\ with major axis in PA=
-52$^{\circ}$.  Since no nuclear emission could be isolated in
the radio, the X-ray map has not been registered.  A gap
between ACIS chips runs through the N lobe and the edge of the S3 chip
is obvious to the SE of the source.}
\label{fig:3c198app}
\end{figure}

\clearpage
\begin{figure}
\includegraphics[keepaspectratio=true,scale=0.80]{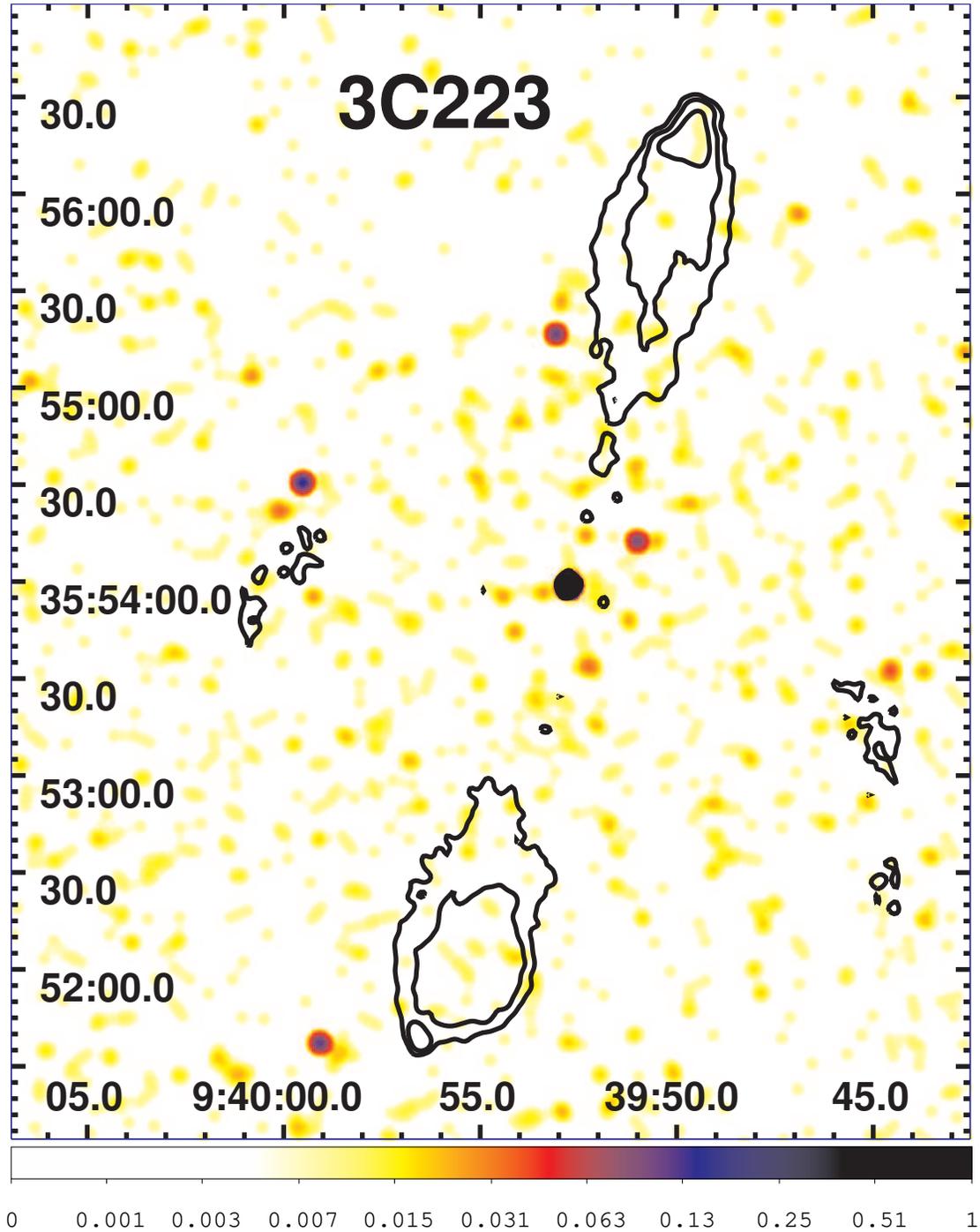}
\caption{The X-ray image of 3C223 for the energy band 0.5-7 keV.  The smoothing
function applied is a Gaussian of FWHM=5.2\arcsec.  The
radio contours (black) come from an 8.4 GHz map kindly supplied by
M. Hardcastle, and start at 0.2 mJy/beam, increasing by factors
of four.  The clean beam is 2.5\arcsec.  The N radio hotspot
falls on chip S2.}
\label{fig:3c223app}
\end{figure}

\clearpage
\begin{figure}
\includegraphics[keepaspectratio=true,scale=0.90]{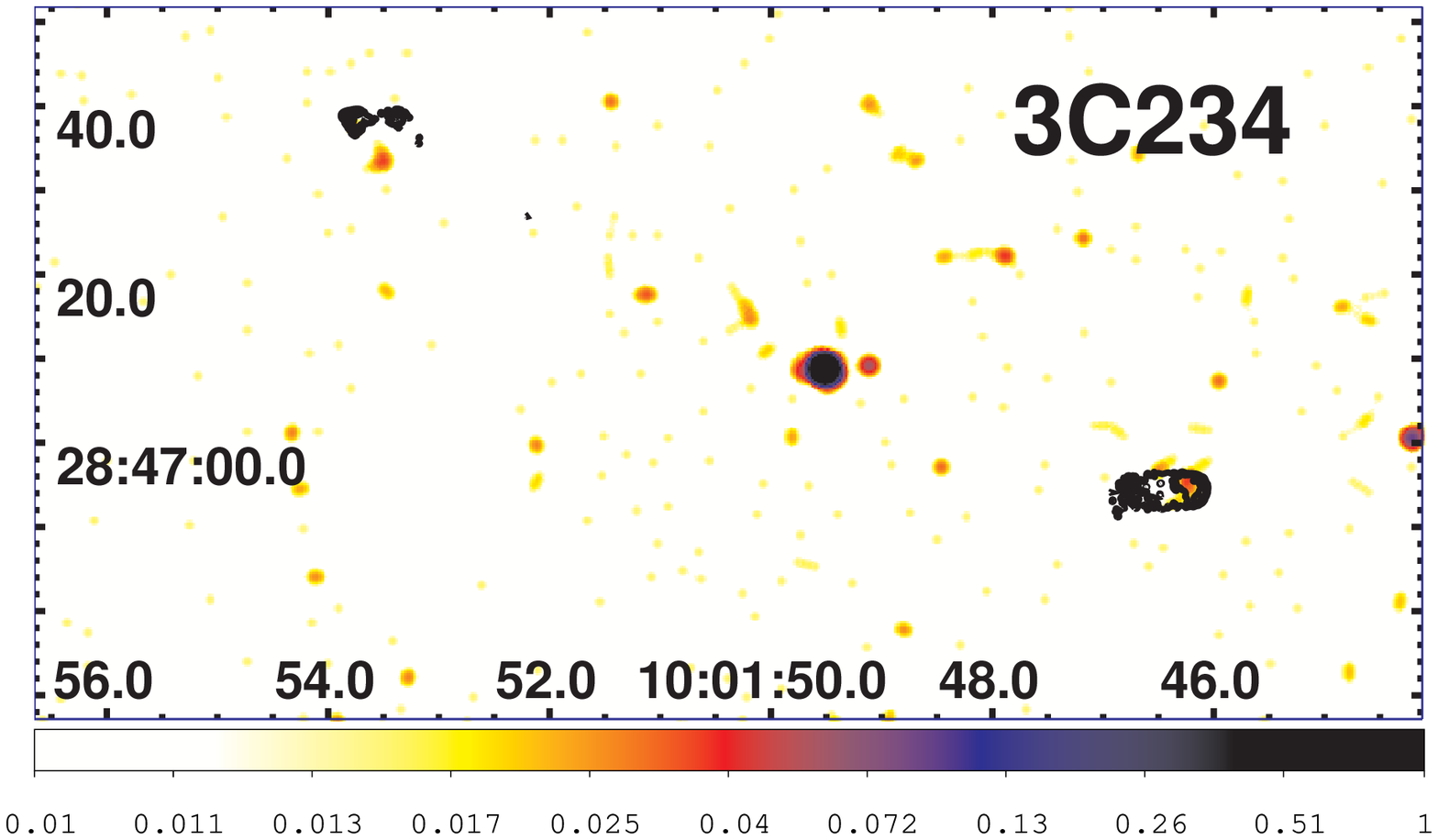}
\caption{The X-ray image of 3C234 for the energy band 0.5-7 keV.  The event
file has been regridded to a pixel size of 0.243\arcsec\ and
smoothed with a Gaussian of FWHM=2.0\arcsec.  The radio
contours (black) come from an 8.4 GHz map kindly supplied by
C. C. Cheung, and start at 0.4 mJy/beam, increasing by factors of
four.  The clean beam is 0.4\arcsec.}
\label{fig:3c234app}
\end{figure}

\clearpage
\begin{figure}
\includegraphics[keepaspectratio=true,scale=0.90]{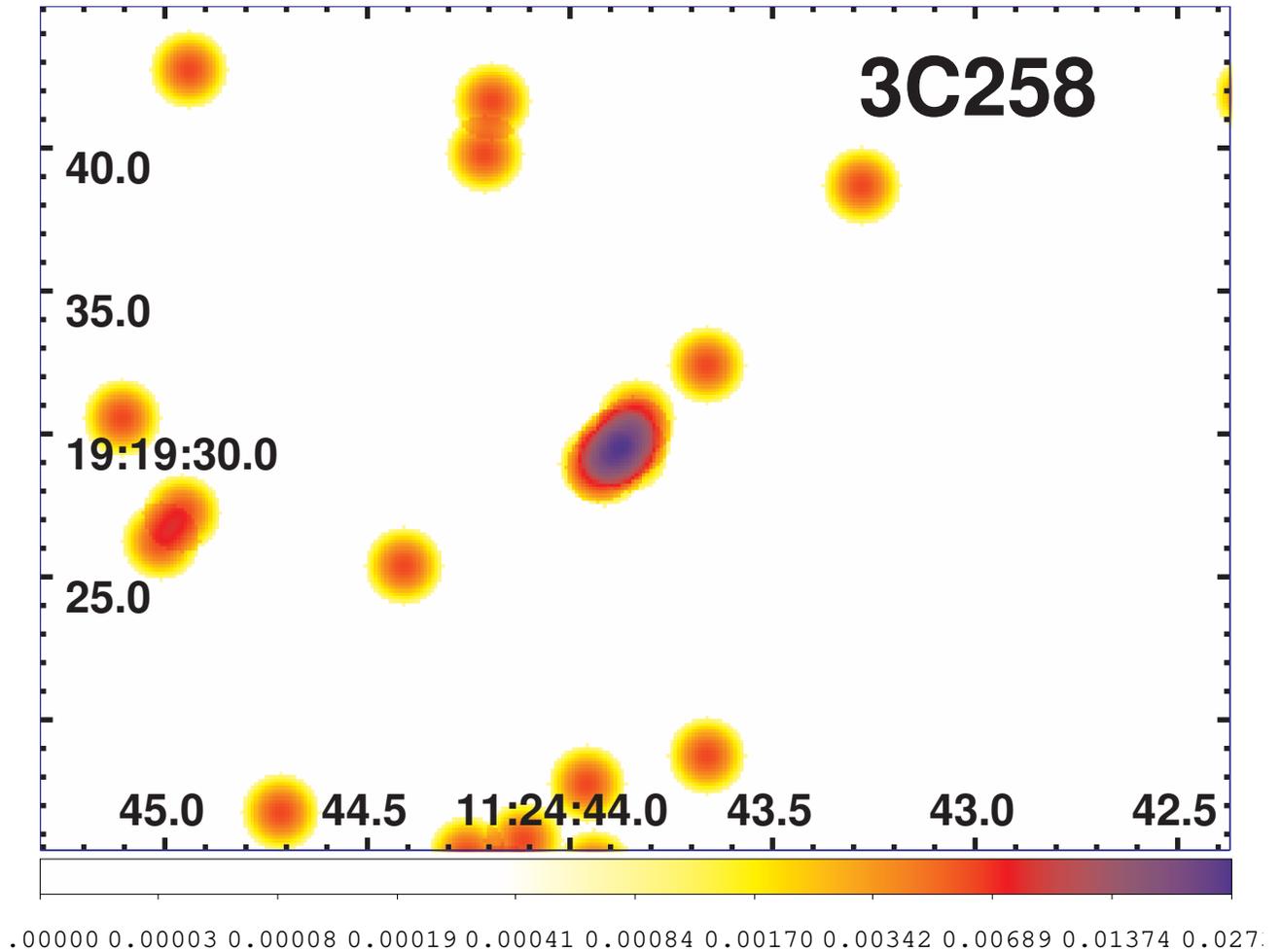}
\caption{The X-ray image of 3C258 for the energy band 0.5-7 keV.  The event
file has been regridded to a pixel size of 0.123\arcsec\ and
smoothed with a Gaussian of FWHM=1.6\arcsec.  There are no radio
contours shown because the total radio intensity consists of a
close (unequal) double with separation of 0.1\arcsec.  See
the 5 GHz MERLIN contours on NED.  Our detection of the nucleus of
the galaxy comprises only 7 counts and is the source at the center of
the figure.}
\label{fig:3c258app}
\end{figure}

\clearpage
\begin{figure}
\includegraphics[keepaspectratio=true,scale=0.90]{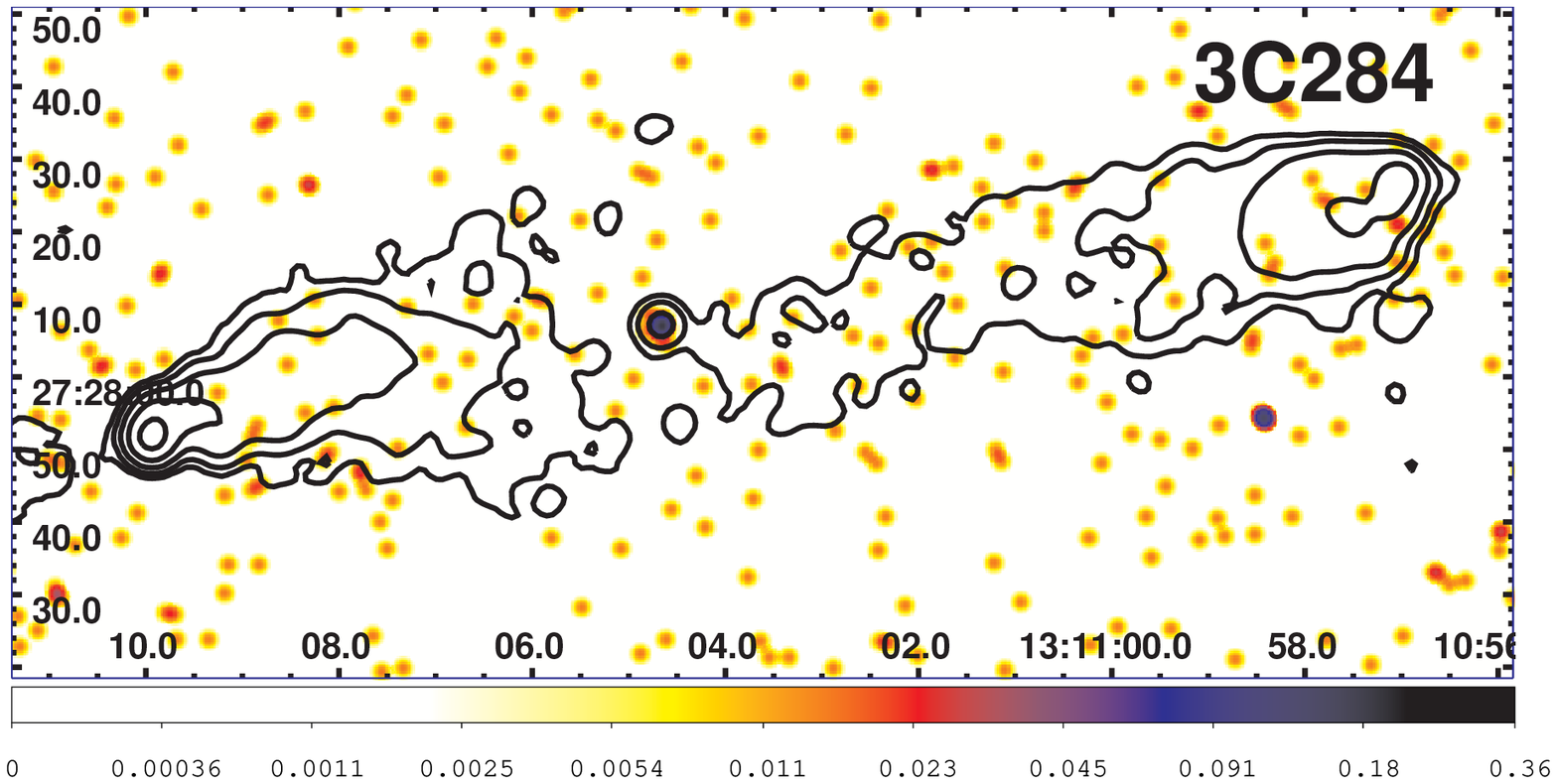}
\caption{The X-ray image of 3C284 for the energy band 0.5-7 keV.  The event
file has been regridded to a pixel size of 0.246\arcsec\ and
smoothed with a Gaussian of FWHM=2.0\arcsec.  The radio
contours (black) come from a 8.0 GHz map kindly supplied by
M. Hardcastle, and start at 0.1 mJy/beam, increasing by factors of
four.  The clean beam is 3.6\arcsec.}
\label{fig:3c284app}
\end{figure}

\clearpage
\begin{figure}
\includegraphics[keepaspectratio=true,scale=0.90]{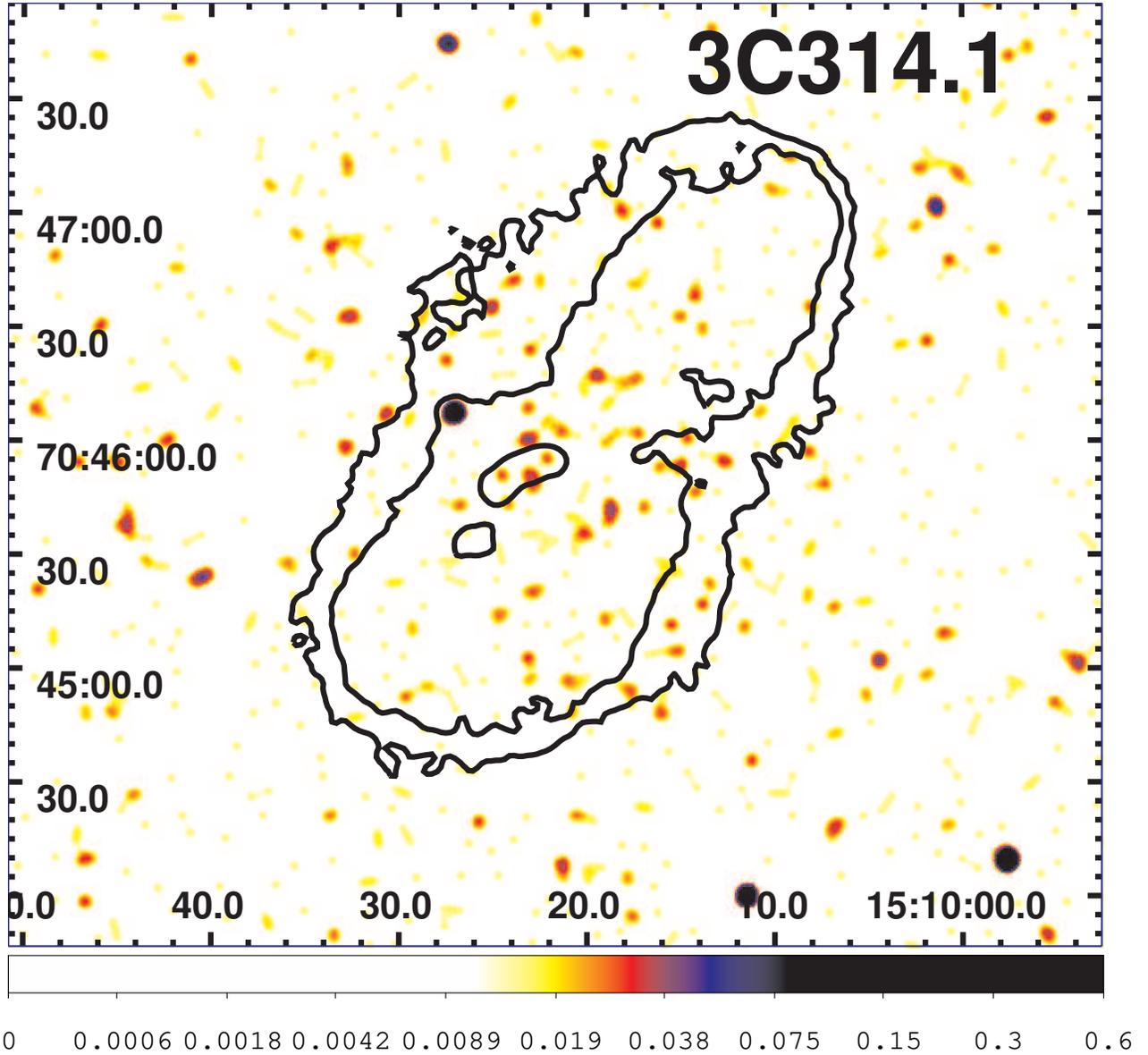}
\caption{The X-ray image of 3C314.1 for the energy band 0.5-7 keV.  The pixel
size is 0.492\arcsec\ and the map has been smoothed with a
Gaussian of FWHM=4.0\arcsec.  The
radio contours (black) come from a 1.5 GHz map downloaded from the
DRAGN website and start at 0.3 mJy/beam, increasing by factors of four.  The
clean beam is 4.2\arcsec.  We used the point source which
lies near the E edge of the radio structure to register the X-ray
image so as to match the radio map.}
\label{fig:3c314.1app}
\end{figure}

\clearpage
\begin{figure}
\includegraphics[keepaspectratio=true,scale=0.90]{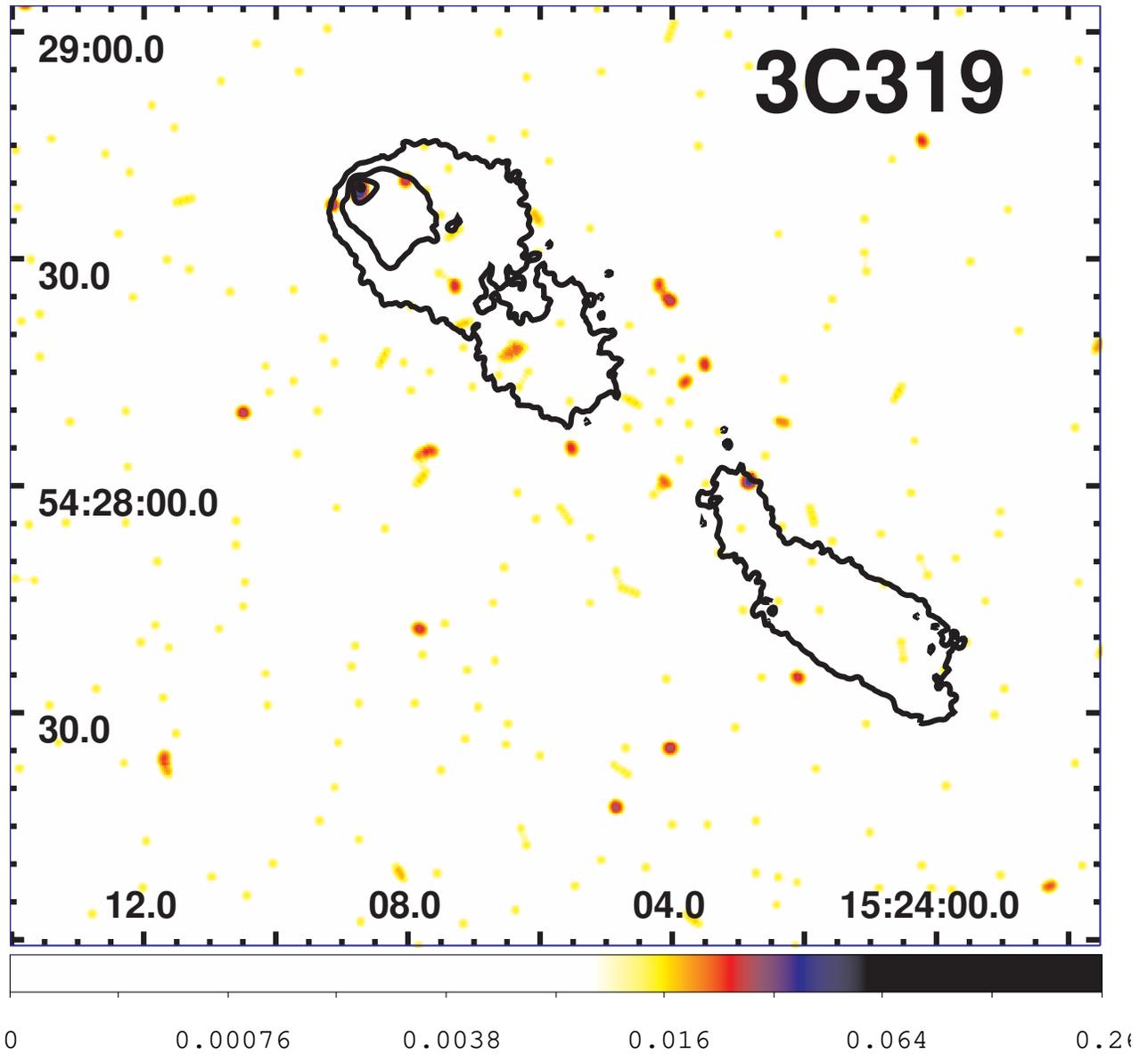}
\caption{The X-ray image of 3C319 for the energy band 0.5-7 keV.  The event
file has been regridded to a pixel size of 0.246\arcsec\ and
smoothed with a Gaussian of FWHM=2.0\arcsec.  The radio
contours (black) come from an 8.4 GHz map kindly provided by
M. Hardcastle, and start at 0.1 mJy/beam, increasing by factors of
four.  The clean beam is 0.9\arcsec.  Since the nucleus has
not been detected in either X-rays or radio, the map is not
registered.  We 'detect' the N radio hotspot with 4 counts.}
\label{fig:3c319app}
\end{figure}

\clearpage
\begin{figure}
\includegraphics[keepaspectratio=true,scale=0.90]{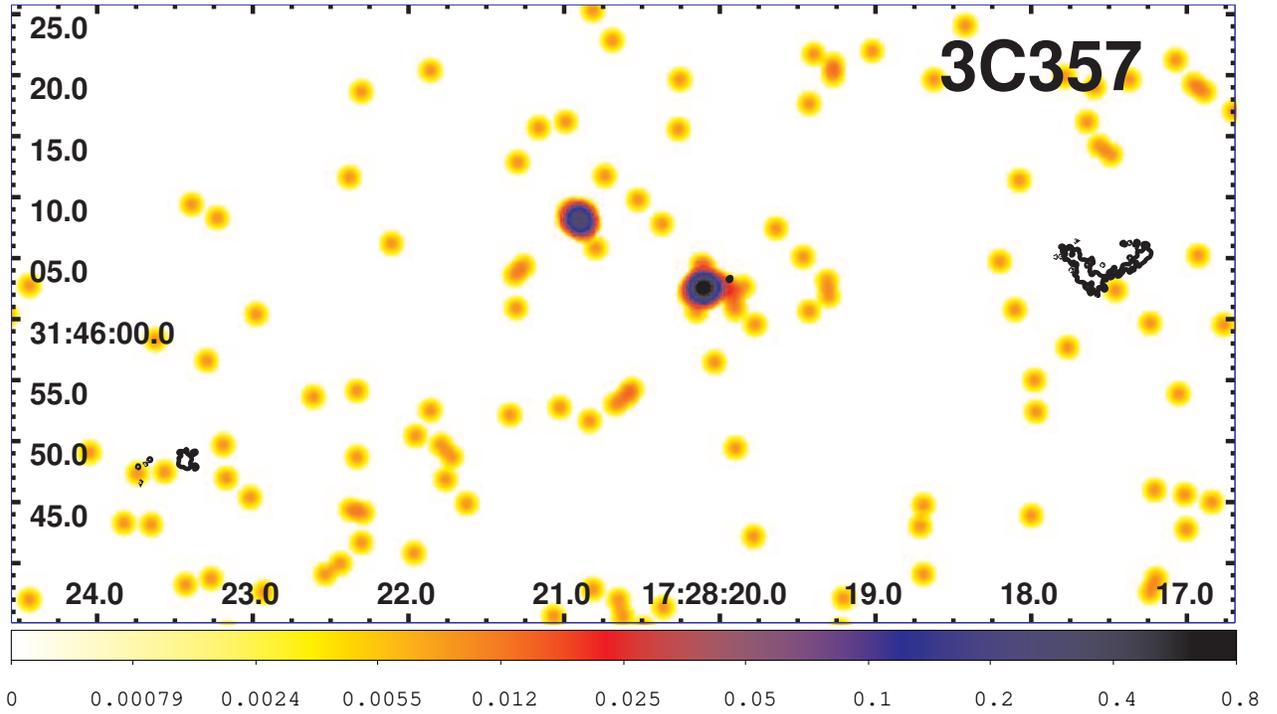}
\caption{The X-ray image of 3C357 for the energy band 0.5-7 keV.  The event
file has been regridded to a pixel size of 0.123\arcsec\ and
smoothed with a Gaussian of FWHM=1.3\arcsec.  The radio
contours (black) come from an 8.4 GHz map kindly supplied by
C. C. Cheung, and start at 0.4 mJy/beam, increasing by factors of
four.  The clean beam is 0.33\arcsec.  The nucleus is the SW
of the pair of bright sources near the field center.}
\label{fig:3c357app}
\end{figure}

\clearpage
\begin{figure}
\includegraphics[keepaspectratio=true,scale=0.80]{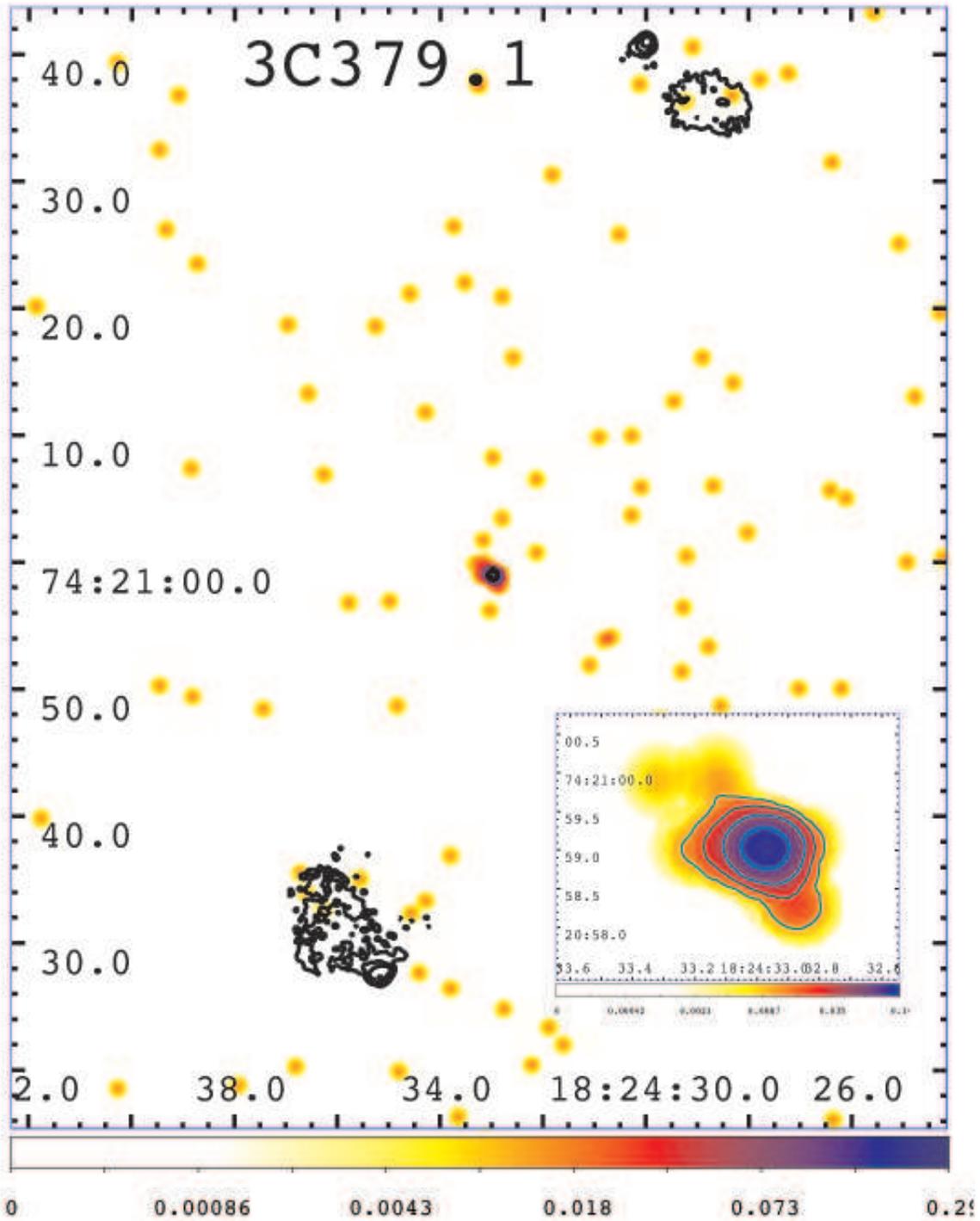}
\caption{The X-ray image of 3C379.1 for the energy band 0.5-7 keV.  The event
file has been regridded to a pixel size of 0.123\arcsec\ and
smoothed with a Gaussian of FWHM=1.0\arcsec.  The radio
contours (black) come from an 8.4 GHz kindly provided by C. C. Cheung,
and start at 0.2 mJy/beam, increasing by factors of four.  The clean
beam is 0.35\arcsec.
{ The 30 counts comprising the nucleus are shown in the insert.  The 
pixel size is 0.0615$^{\prime\prime}$ and the smoothing function has a 
FWHM of 0.65$^{\prime\prime}$.  X-ray contours begin at 0.012
cnts/pixel and increase by factors of two.}
}
\label{fig:3c379.1app}
\end{figure}

\clearpage
\begin{figure}
\includegraphics[keepaspectratio=true,scale=0.90]{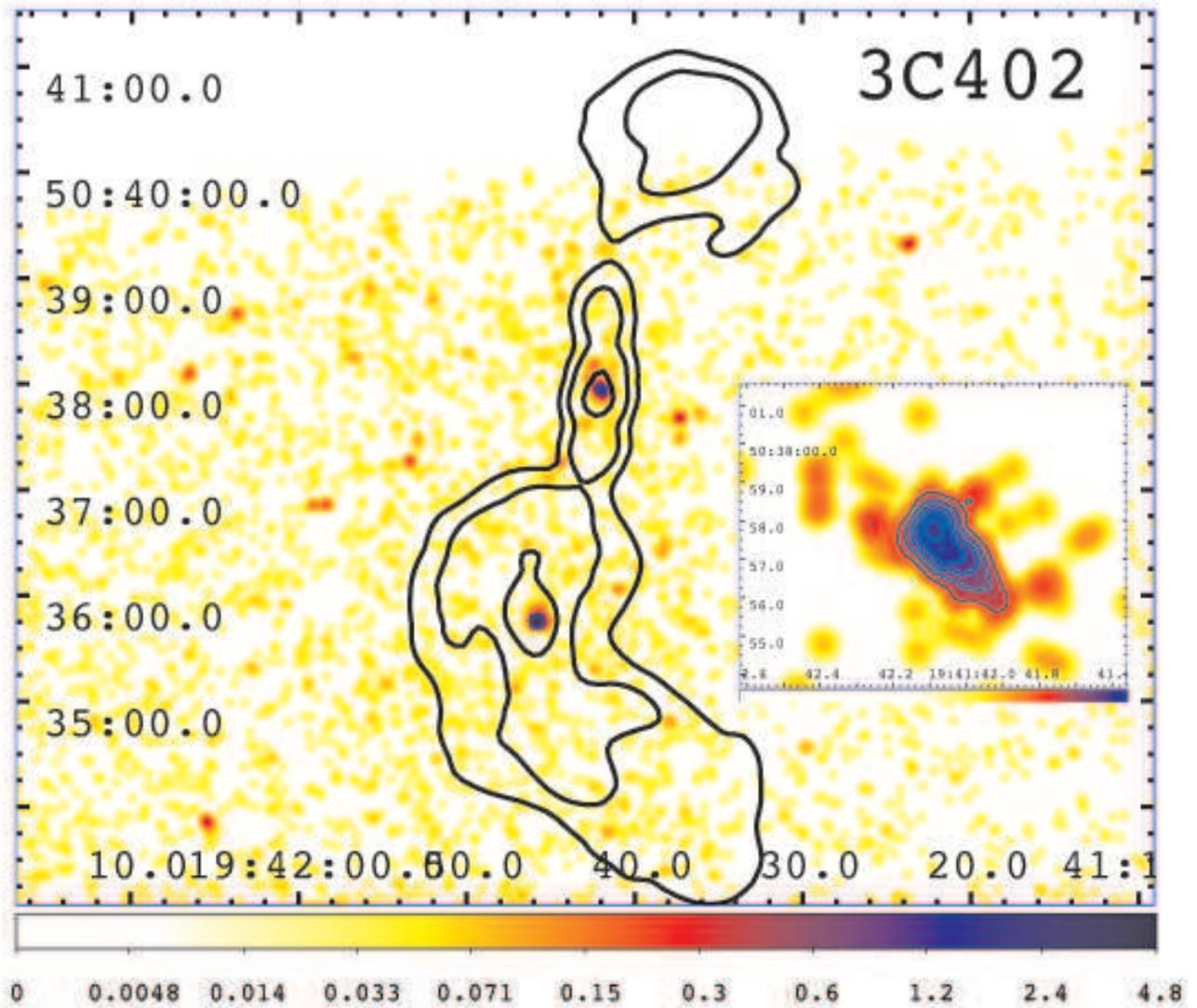}
\caption{The X-ray image of 3C402 for the energy band 0.5-7 keV.  The event
file has been regridded to a pixel size of 0.984\arcsec\ and
smoothed with a Gaussian of FWHM=5.8\arcsec.  The radio
contours (black) come from a 1.5 GHz map downloaded from the NVAS and
start at 4 mJy/beam, increasing by factors of four.  The clean beam is
19.7\arcsec\ x 13\arcsec\ with major axis in PA=
90$^{\circ}$.  There are two bright galaxies, each of which is 
detected in X-rays, and both appear to contribute to the complex 
radio morphology.  The N radio lobe extends off the ACIS S3 chip.
Since neither nucleus (both in radio and X-rays) is point-like,
the registration of the X-ray image is not as accurate as usual.
{ An X-ray image of the N nucleus is shown in the insert.  The
pixel size is 0.0615$^{\prime\prime}$ and the smoothing function has
FWHM=0.8$^{\prime\prime}$.  Contours are linear: 0.015, 0.025,
... 0.055 cnts/pix.  The structure defined by the contours contains
about 50 counts.}}
\label{fig:3c402app}
\end{figure}

\clearpage
\begin{figure}
\includegraphics[keepaspectratio=true,scale=0.90]{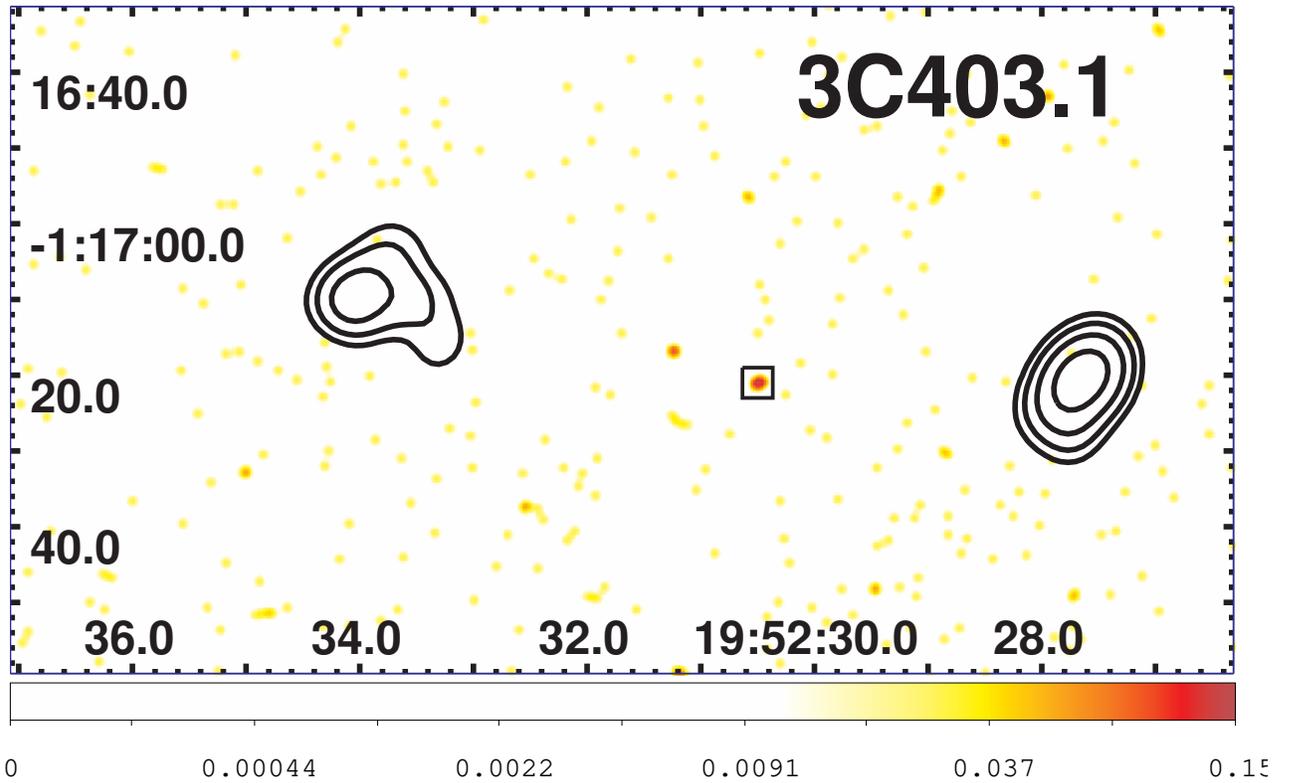}
\caption{The X-ray image of 3C403.1 for the energy band 0.5-7 keV.  The event
file has been regridded to a pixel size of 0.246\arcsec\ and
smoothed with a Gaussian of FWHM=1.4\arcsec.  The radio
contours (black) come from a 0.3 GHz map and start at 16 mJy/beam,
increasing by factors of two.  The clean beam is
7.0\arcsec\ x 6.2\arcsec\ with major axis in PA=-20
$^{\circ}$.  The small box indicates the NED position and we find
a weak source there.}
\label{fig:3c403.1app}
\end{figure}

\clearpage
\begin{figure}
\includegraphics[keepaspectratio=true,scale=0.90]{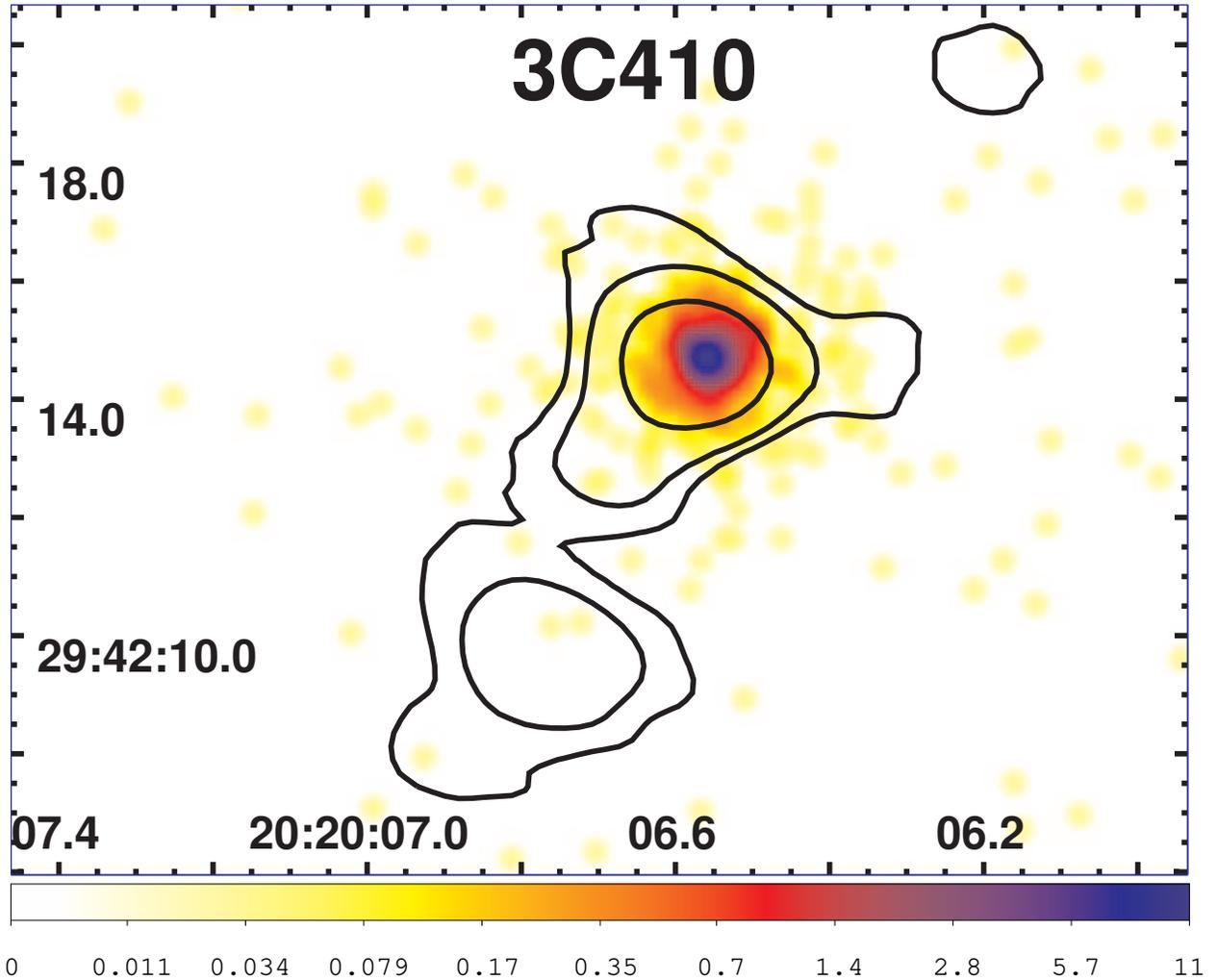}
\caption{The X-ray image of 3C410 for the energy band 0.5-7 keV.  The event
file has been regridded to a pixel size of 0.0615\arcsec\ and
smoothed with a Gaussian of FWHM=0.36\arcsec.  The radio
contours (black) come from a 43 GHz map downloaded from the NVAS and
start at 6 mJy/beam, increasing by factors of four.  The clean beam is
1.7\arcsec.  The shift in the X-ray map required to align
the X-ray source with the central component of a 5 GHz map is
1\arcsec\ in declination; far larger than normal.}
\label{fig:3c410app}
\end{figure}

\clearpage
\begin{figure}
\includegraphics[keepaspectratio=true,scale=0.90]{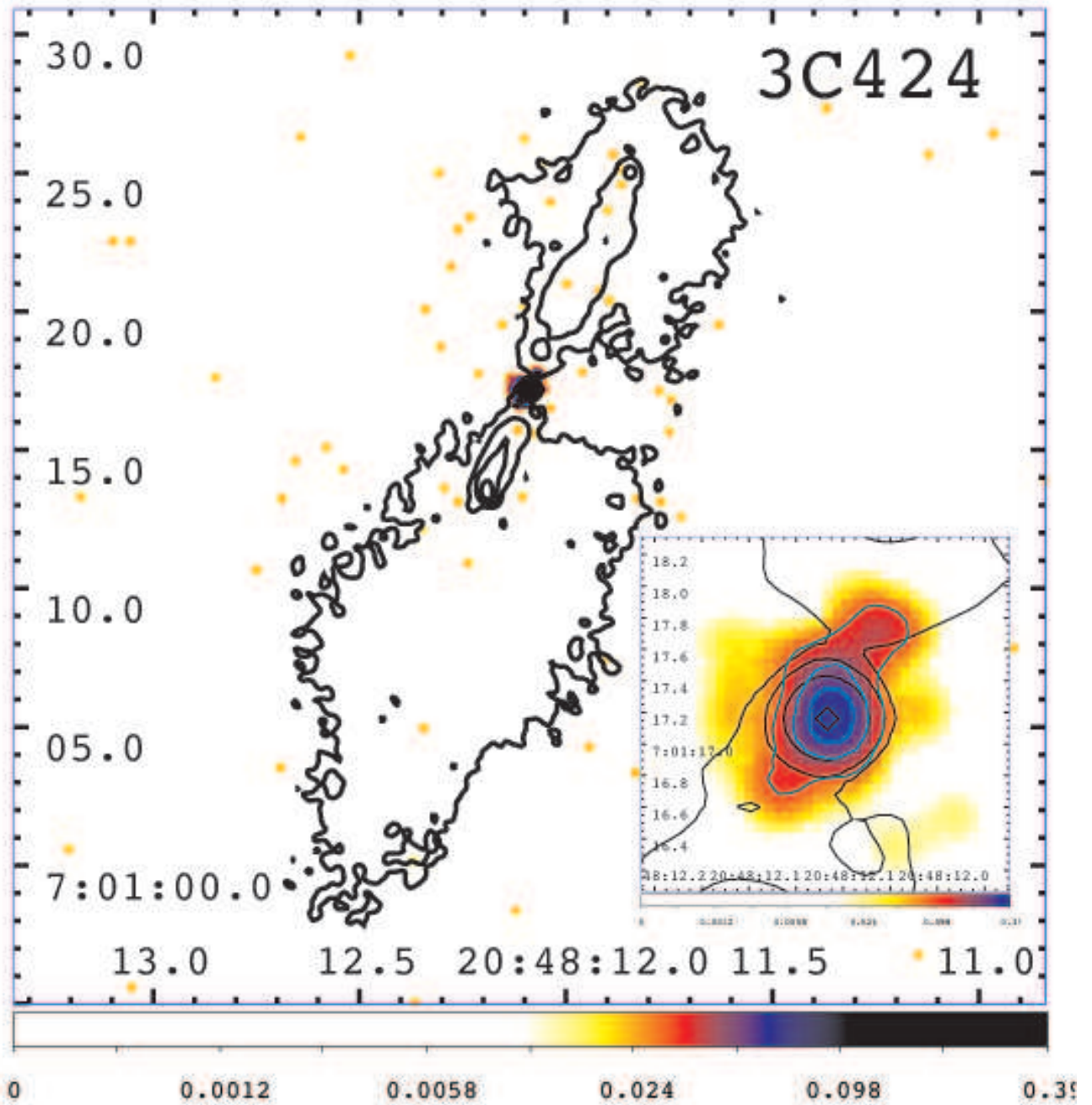}
\caption{The X-ray image of 3C424 for the energy band 0.5-7 keV.  The event
file has been regridded to a pixel size of 0.0615\arcsec\ and
smoothed with a Gaussian of FWHM=0.36\arcsec.  The radio
contours (black) come from a 8.5 GHz map kindly provided by
M. Hardcastle, and start at 0.05 mJy/beam, increasing by factors of
four.  The clean beam is 0.25\arcsec.
{ The insert shows a blowup of the nucleus.  Countours are linear: 0.1,
0.2, 0.3 and 0.4 cnts/pix.  The smoothing function is a Gaussian of
FWHM=0.36$^{\prime\prime}$}}
\label{fig:3c424app}
\end{figure}

\clearpage
\begin{figure}
\includegraphics[keepaspectratio=true,scale=0.90]{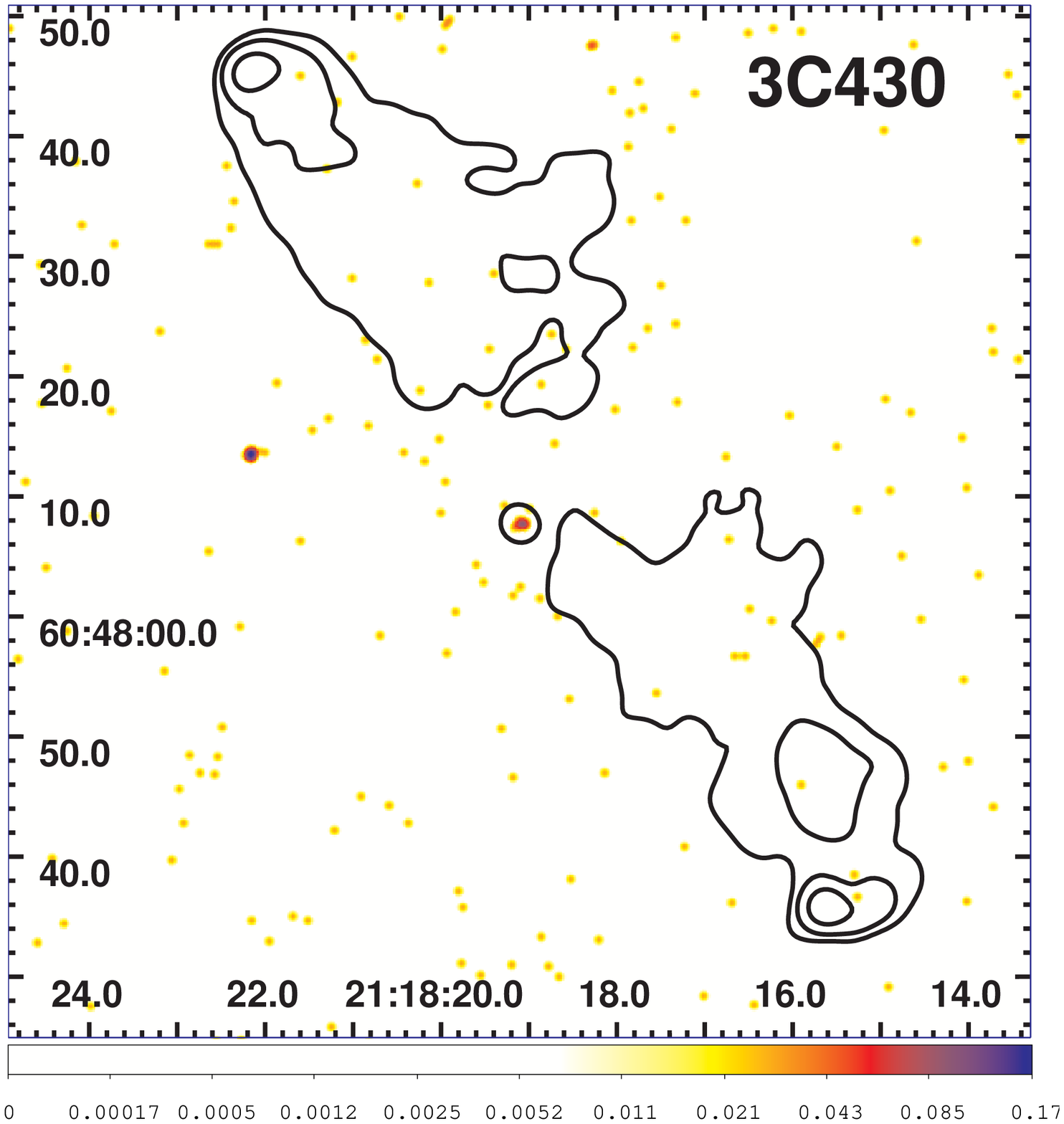}
\caption{The X-ray image of 3C430 for the energy band 0.5-7 keV.  The event
file has been regridded to a pixel size of 0.123\arcsec\ and
smoothed with a Gaussian of FWHM=0.7\arcsec.  The radio
contours (black) come from a 4.9 GHz map downloaded from the NVAS and
start at 1 mJy/beam, increasing by factors of four.  The clean beam
is 1.3\arcsec.}
\label{fig:3c430app}
\end{figure}

\clearpage
\begin{figure}
\includegraphics[keepaspectratio=true,scale=0.90]{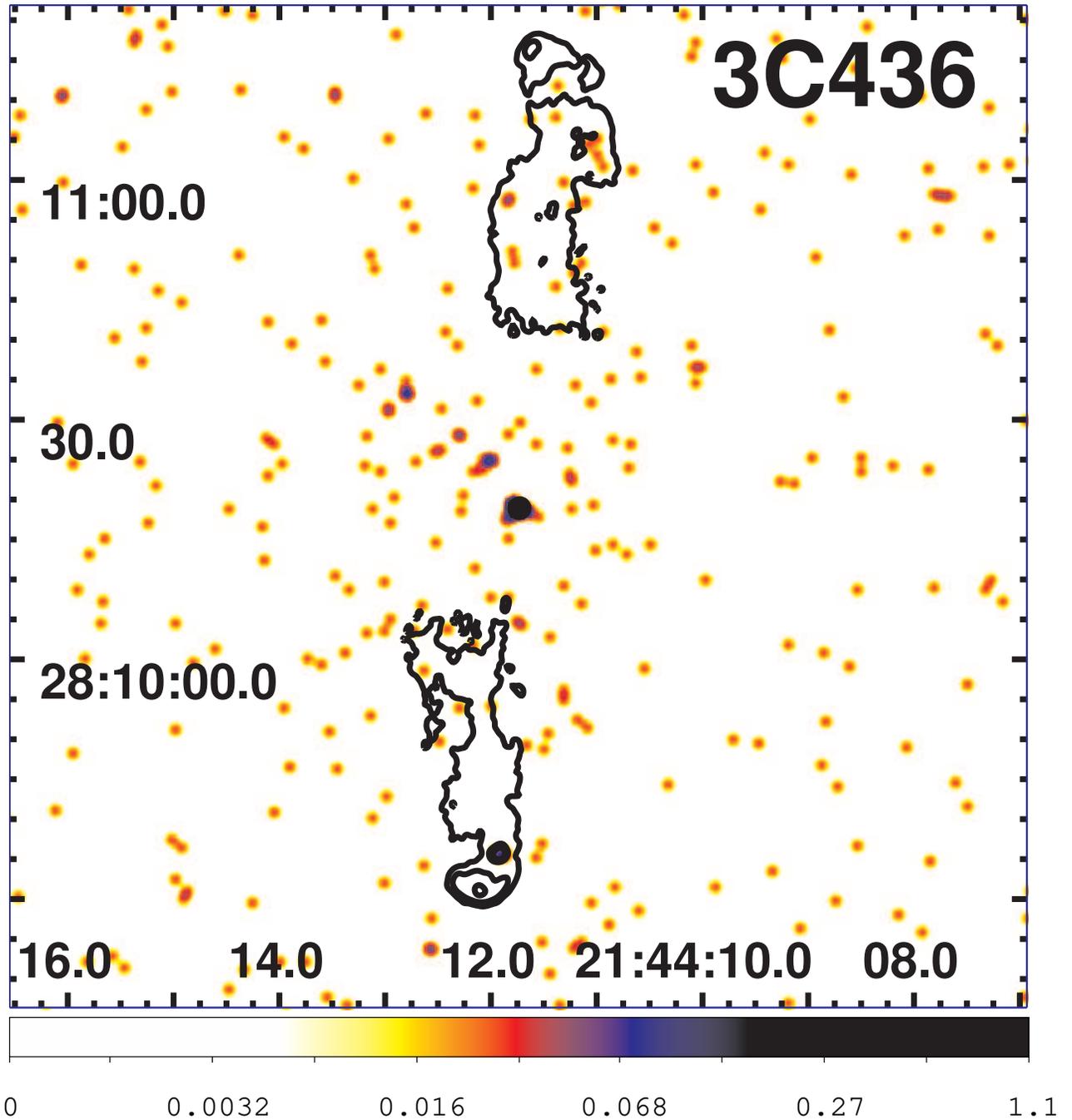}
\caption{The X-ray image of 3C436 for the energy band 0.5-7 keV.  The event
file has been regridded to a pixel size of 0.246\arcsec\ and
smoothed with a Gaussian of FWHM=1.4\arcsec.  The radio
contours (black) come from an 8.4 GHz map kindly supplied by
M. Hardcastle, and start at 0.2 mJy/beam, increasing by factors of
four.  The clean beam is 0.75\arcsec.  The hotspot near the
tip of the S lobe is detected with 4 counts.}
\label{fig:3c436app}
\end{figure}

\clearpage
\begin{figure}
\includegraphics[keepaspectratio=true,scale=0.90]{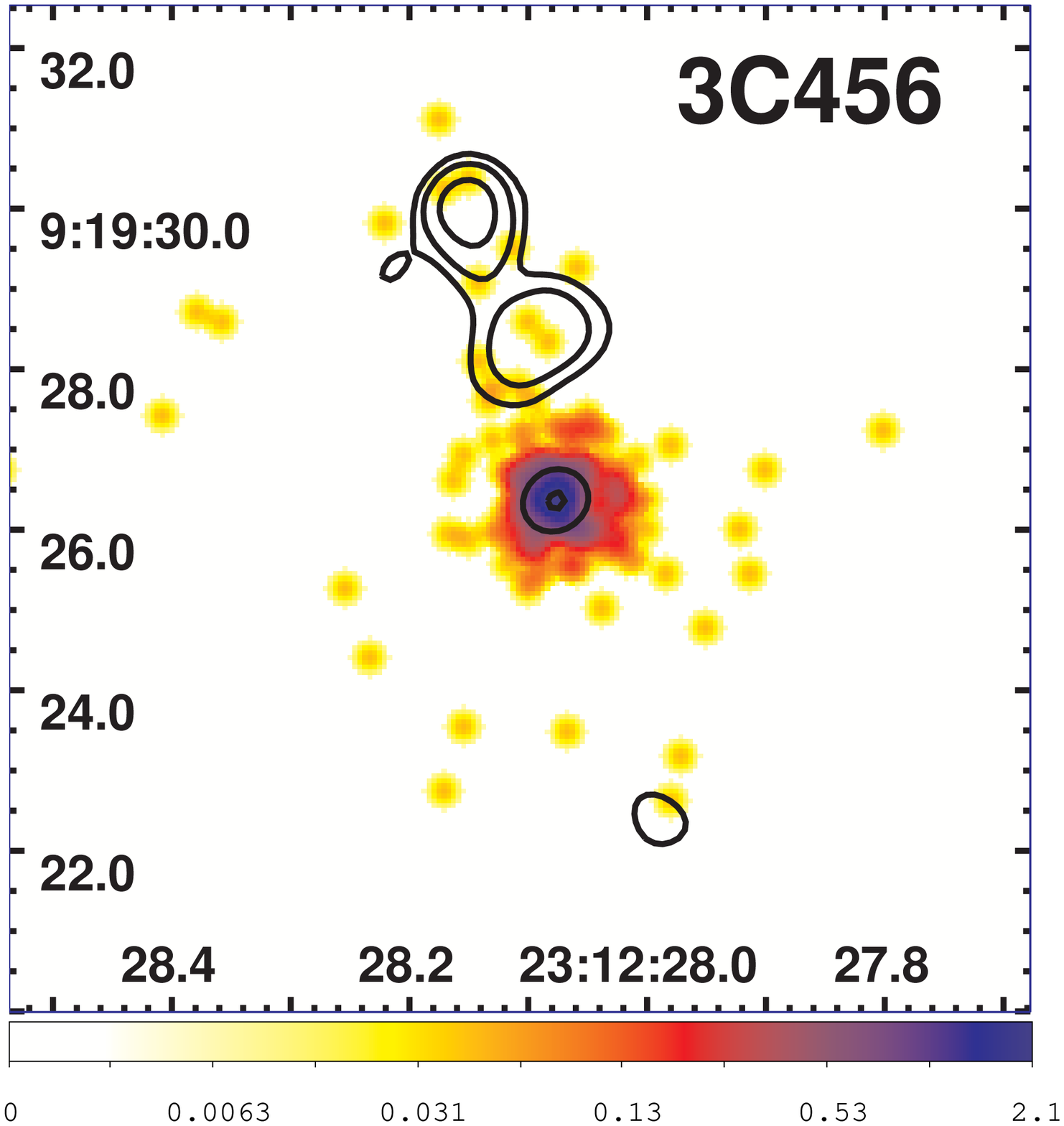}
\caption{The X-ray image of 3C456 for the energy band 0.5-7 keV.  The event
file has been regridded to a pixel size of 0.0615\arcsec\ and
smoothed with a Gaussian of FWHM=0.3\arcsec.  The radio
contours (black) come from a 4.9 GHz map kindly supplied by
M. Hardcastle, and start at 5 mJy/beam, increasing by factors of
four.  The clean beam is 0.5\arcsec.}
\label{fig:3c456app}
\end{figure}

\clearpage
\begin{figure}
\includegraphics[keepaspectratio=true,scale=0.90]{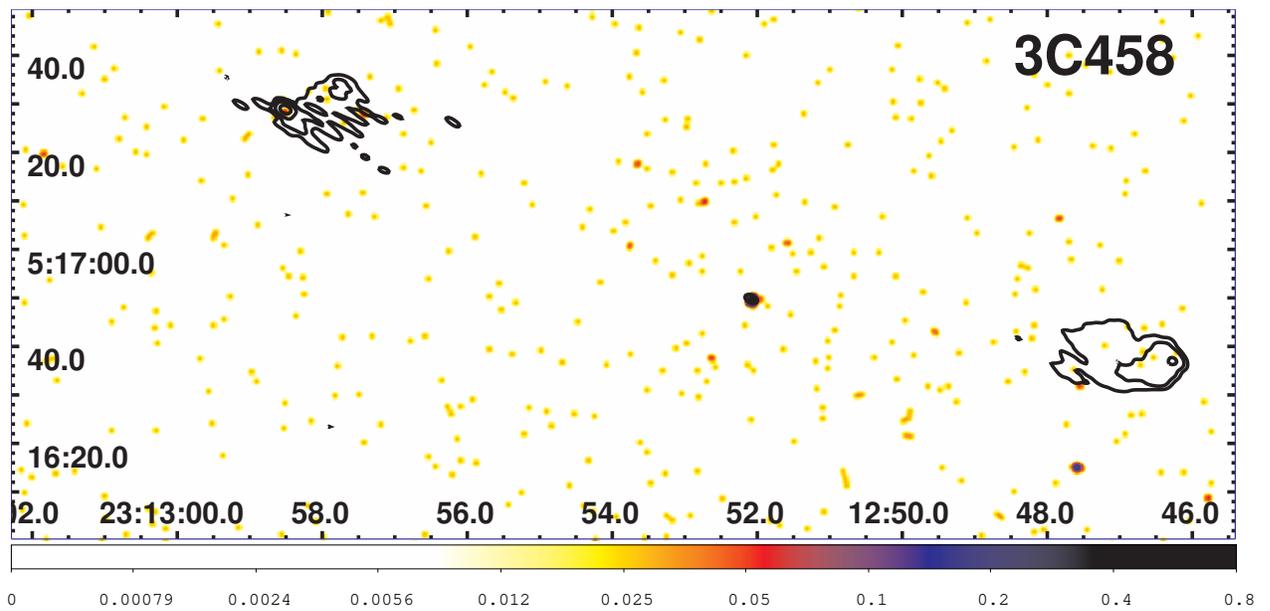}
\caption{The X-ray image of 3C458 for the energy band 0.5-7 keV.  The event
file has been regridded to a pixel size of 0.246\arcsec\ and
smoothed with a Gaussian of FWHM=1.5\arcsec.  The radio
contours (black) come from a 4.9 GHz map downloaded from the NVAS and
start at 1 mJy/beam, increasing by factors of four.  The clean beam
is 2.3\arcsec\ x 1.4\arcsec\ with major axis in PA=65
$^{\circ}$.  We detect the NE hotspot with 4 counts.}
\label{fig:3c458app}
\end{figure}

\clearpage
\begin{figure}
\includegraphics[keepaspectratio=true,scale=0.90]{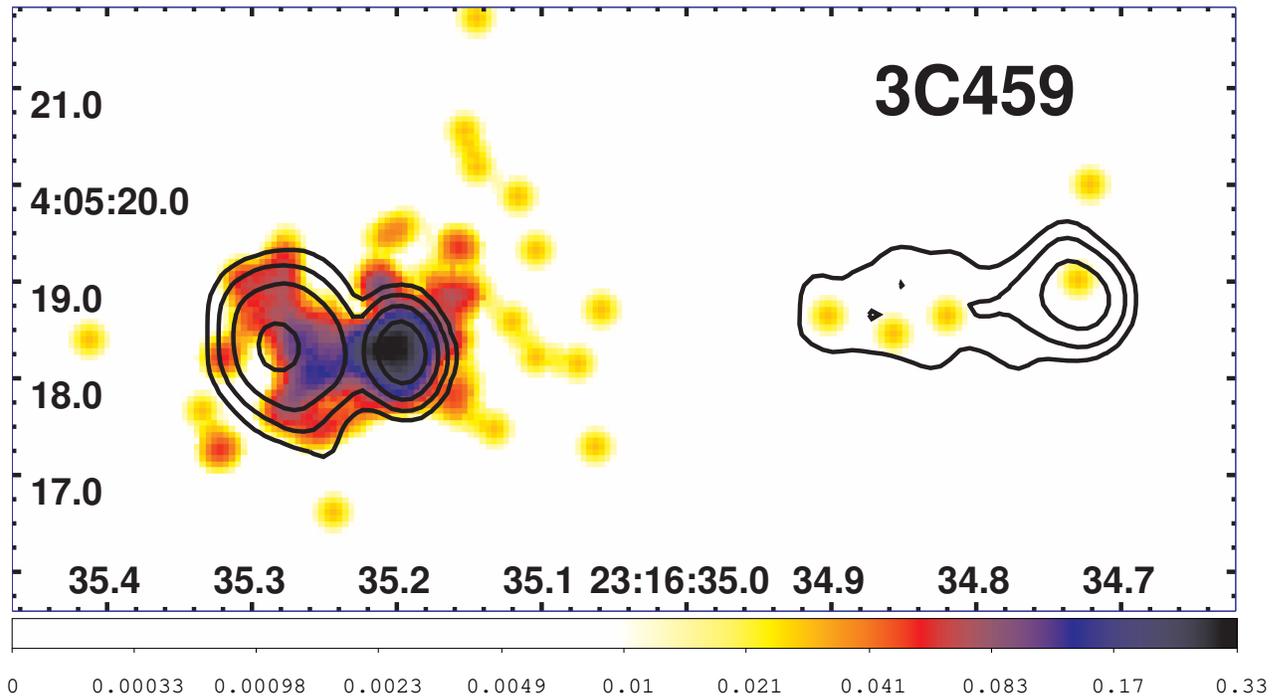}
\caption{The X-ray image of 3C459 for the energy band 0.5-7 keV.  The event
file has been regridded to a pixel size of 0.0615\arcsec\ and
smoothed with a Gaussian of FWHM=0.36\arcsec.  The radio
contours (black) come from a 4.9 GHz map downloaded from the NVAS and
start at 2 mJy/beam, increasing by factors of four.  The clean beam
is 0.5\arcsec\ x 0.4\arcsec\ with major axis in PA=9
$^{\circ}$.}
\label{fig:3c459app}
\end{figure}

\end{document}